\documentclass[%
twocolumn, 
notitlepage, 
aps,
prb, 
10pt, 
superscriptaddress,
floatfix,
letterpaper,
reprint,
bibnotes,
amsmath,amssymb
]{revtex4-2}

\usepackage{hyperref}
\usepackage{graphicx}
\usepackage{dcolumn}
\usepackage{bm}
\usepackage[compat=1.1.0]{tikz-feynman}
\usepackage{physics}
\usepackage{lipsum}
\usepackage[ruled,vlined]{algorithm2e}

\definecolor{linkcolor}{HTML}{399B03}
\definecolor{urlcolor}{HTML}{399B03}
\hypersetup{pdfstartview=FitH, linkcolor=linkcolor,urlcolor=urlcolor,colorlinks=true}
\hypersetup{
    unicode=false,          
    pdftoolbar=true,        
    pdfmenubar=true,        
    pdffitwindow=false,     
    pdfstartview={FitH},    
    pdfauthor={},         
    colorlinks=true,        
    linkcolor=blue,         
    citecolor=red,          
}

\begin{document}

\title{Fully Self-Consistent Finite-Temperature $GW$ in Gaussian Bloch Orbitals for Solids}

\author{Chia-Nan Yeh}
\affiliation{%
    Department of Physics, University of Michigan, Ann Arbor, Michigan 48109, USA
}%

\author{Sergei Iskakov}
\affiliation{%
    Department of Physics, University of Michigan, Ann Arbor, Michigan 48109, USA
}%

\author{Dominika Zgid}%
\affiliation{%
Department of Chemistry, University of Michigan, Ann Arbor, Michigan 48109, USA
}%
\affiliation{%
Department of Physics, University of Michigan, Ann Arbor, Michigan 48109, USA
}%

\author{Emanuel Gull}%
\affiliation{%
 Department of Physics, University of Michigan, Ann Arbor, Michigan 48109, USA
}%

\date{\today}

\begin{abstract}
We present algorithmic and implementation details for the fully self-consistent finite-temperature $GW$ method in Gaussian Bloch orbitals for solids. 
Our implementation is based on the finite-temperature Green's function formalism in which all equations are solved on the imaginary axis, without resorting to analytical continuation during the self-consistency. 
No quasiparticle approximation is employed and all matrix elements of the self-energy are explicitly evaluated. 
The method is tested by evaluating the band gaps of selected semiconductors and insulators. 
We show agreement with other, differently formulated finite-temperature sc$GW$ implementations when finite-size corrections and basis set errors are taken into account.
By migrating computationally intensive calculations to GPUs, we obtain scalable results on large supercomputers with nearly optimal performance. 
Our work demonstrates the applicability of Gaussian orbital based sc$GW$ for $\emph{ab initio}$ correlated materials simulations and provides 
a sound starting point for embedding methods built on top of $GW$.
\end{abstract}

\maketitle

\section{Introduction}
The $GW$ method~\cite{Hedin65} provides direct access to single-particle excitation spectra, unlike ground-state methods such as the density functional theory (DFT)~\cite{KS_DFT_1965}. 
\emph{Ab initio} simulations of single particle excitation spectra are essential for direct comparison to experiment such as angle-resolved photoemission spectroscopy (ARPES).  
$GW$ has been widely applied to weakly correlated systems, such as semiconductors~\cite{Aryasetiawan_GW_review_1998,GW_review_Reining_Rubio_2002}. Due to the increasing availability of computing power, it is rapidly becoming an alternative to DFT~\cite{Aryasetiawan_GW_review_1998,GW_review_Reining_Rubio_2002}.
The $GW$ method also frequently serves as the first step in embedding frameworks designed to include strong correlations~\cite{DMFT_RMP_1996,GW_EDMFT_Sun_2002,GW_EDMFT_Biermann_2003,electronic_strcuture_w_DMFT_RMP_2006,Dominika_SEET_2017,Alexie_SEET_PRB2015,SEET_NiO_MnO_2020,SEET_perovskites_Yeh_2021,SEET_GFCCSD_Yeh_2021,Nilsson17,GW_EDMFT_PRM_Philipp20,Boehnke16,full_cell_dmft_chan}. 

In Hedin's seminal paper~\cite{Hedin65} a set of exact self-consistent equations was introduced, describing an expansion of the self-energy in terms of the screened interaction. 
Such an expansion is especially beneficial for metallic systems, where expansions in terms of the bare (i.e. un-screened) interactions may diverge. 
Hedin's equations are relations between the Green's function, self-energy, vertex function, polarization function, and screened Coulomb interaction. 
In the so-called $GW$ approximation, Hedin's equations are truncated to the first order in the screened Coulomb interaction. 
The self-consistent solution of these equations is guaranteed to satisfy certain conservation laws and is thermodynamically consistent~\cite{Baym_Kadanoff_1961,Baym_1962,Hedin65}. 

Despite its theoretical simplicity, obtaining results from the $GW$ approximation is orders of magnitude more expensive than solving the equations of DFT, and implementations of the  $GW$ method for materials have not yet reached the maturity of DFT codes, where consistent results for both molecules and solids can be obtained from independent codes with different numerical setups~\cite{VASP_1996,QE_2009,CRYSTAL_2018,PySCF_2020,CP2K_2020,wien2k_2020}. 
A one-to-one comparison of $GW$ implementations is complicated by a dependence on basis sets (e.g. plane waves, linearized augmented plane waves (LAPW), Gaussian orbitals (GTO)), differences in correcting for finite size effects, differences in methodologies for evaluating quasiparticle peaks and band gaps, and the effect of additional approximations beyond the truncation to first order in the screened interaction. These approximations may consist of a combination of `numerical' approximations (such as the choice of finite basis set in space or frequency, or analytic continuation) and  `theoretical' approximations. 
For molecular systems, multiple comparisons of results from different implementations of the one-shot variant of $GW$ ($G_{0}W_{0}$) have been published before~\cite{GW100_benchmark,GW100_benchmark2,GW100_benchmark3,Bruneval2021}. 

Common additional theoretical approximations introduced in practical $GW$ calculations include approximations to the self-consistency as well as quasiparticle approximations. 
The $GW$ method is frequently executed non-selfconsistently. This approximation is referred to as  $G_{0}W_{0}$~\cite{G0W0_Louie_1985,G0W0_Louie_1986,G0W0_Sham_1988,G0W0_SPEX_2010,G0W0_VASP_2014,West_GW_Govoni_2015,CP2K_G0W0_2016,G0W0_solids_CP2K_Wilhelm_2017,LowScaling_G0W0_CP2K_2018,G0W0_truncated_C_GDF_CP2K_2021,G0W0_Zhu2021}. 
Consequently, $G_{0}W_{0}$ results depend on the choice of the starting single-particle Green's function, $G_0$. 
The selection of a proper starting point requires empirical knowledge of the system of interest.
For the band gaps of semiconductors, common choices of $G_0$ include the DFT Green's function with LDA and PBE functionals~\cite{G0W0_Louie_1985,G0W0_Louie_1986,G0W0_Sham_1988,G0W0_SPEX_2010,G0W0_VASP_2014,West_GW_Govoni_2015,CP2K_G0W0_2016,G0W0_solids_CP2K_Wilhelm_2017,LowScaling_G0W0_CP2K_2018,G0W0_truncated_C_GDF_CP2K_2021,G0W0_Zhu2021}. 
$G_{0}W_{0}$ calculations also employ the quasiparticle approximation, which approximates the frequency dependence of the self-energy by solving the quasiparticle equation in the Kohn-Sham orbital basis. 
Significant improvement over DFT results has been observed due to well-defined single-particle excitations in $GW$. 
When compared to experimental data, the somewhat fortuitous agreement has been attributed to an error cancellation between the lack of self-consistency and the absence of vertex corrections in the self-energy and the polarizability~\cite{GW_offset_gamma_Kotani2007,GW_vertex_Gruneis_2014,scGW_w_vertex_Andrey_2016,scGW_also_vertex_Andrey_2017}. 

Eigenvalue self-consistent $GW$~\cite{evGW_Louie_1991,evGW_Louie_1994,evGW_Kresse_2007} and quasiparticle self-consistent $GW$ (QS$GW$)~\cite{QSGW_Sergey_2004,QSGW_Sergey_2006,QSGW_Fabien_2006,QSGW_w_vertex_Kresse_2007} attempt to eliminate the starting-point dependence with different levels of self-consistency. 
In QS$GW$, an effective non-local static potential is self-consistently determined in the presence of the $GW$ self-energy to construct an optimal one-body reference Hamiltonian~\cite{QSGW_Sergey_2006}. 
While independent of the starting solution, QS$GW$ still employs the quasiparticle approximation and obtains the self-energy only at certain frequency points. 

The solution of the fully self-consistent $GW$ (sc$GW$) approximation became feasible only in recent years, due to numerous numerical advancements, such as the representation of dynamical quantities~\cite{Legendre_Boehnke_2011,ImaginaryTime_LaplaceTrans_Kaltak2014,IR_Hiroshi_2017,Chebyshev_Gull_2018,Legendre_Dong2020,Minimax_Kaltak_2020,sparse_sampling_Jia_2020,DLR_Kaye_2021} and low-scaling optimizations that exploit the locality of the self-energy~\cite{cubicGW_spatialFFT_VASP_Kaltak2014,cubicGW_Peitao_VASP_2016,linear_scGW_Kutepove_2021}. 
Several fully self-consistent $GW$ implementations have been reported for molecules~\cite{scGW_molecules_2009,scGW_finite_systems_2012,scGW_molecules_2014} and periodic systems~\cite{scGW_KandSi_1998,relativistic_scGW_Andrey_2012,scGW_also_vertex_Andrey_2017, scGW_VASP_2018}. 
Nevertheless, reaching agreement between different sc$GW$ implementations remains challenging~\cite{scGW_also_vertex_Andrey_2017, scGW_VASP_2018}. 
Due to the correlated nature of the $GW$ approximation, both the core-valence interactions and the interactions between occupied and unoccupied states are included beyond a single-particle picture. 
Therefore, the quality of the basis sets for unoccupied states and the treatment of core electrons becomes more important than in DFT~\cite{GW_LAPW_nlo,AgX_LAPW_HLOs,GW_core_valence_interactions}. 

In this paper, we present a formulation of fully self-consistent finite-temperature $GW$ in Gaussian Bloch orbitals for solids.
The method is based on the finite-temperature Green's function formalism on the imaginary axis and thus does not require analytical continuation during the self-consistent loop. 
No quasiparticle approximation is employed. 
It relies on sparse sampling on the imaginary axis~\cite{sparse_sampling_Jia_2020} using the intermediate representation~\cite{IR_Hiroshi_2017}, Gaussian density fitting for the decomposition of the bare Coulomb integrals~\cite{RI_S_ref1_1993,RI_S_ref2_2016,RI_S_ref3_2017}, and Nevanlinna analytical continuation~\cite{Nevanlinna_Jiani_2021} to extract data on the real axis from Green's functions evaluated on the imaginary axis. 
The implementation makes efficient use of parallel GPU architectures. 

Extensions of the present work in the presence of strong electron correlations~\cite{SEET_NiO_MnO_2020,SEET_GFCCSD_Yeh_2021,SEET_perovskites_Yeh_2021} or relativistic effects~\cite{scGW_X2C1e_Yeh} have been discussed previously without presenting details of the $GW$ implementation. 
Here, we focus on implementation details and discuss finite-size effects and finite-size convergence, convergence with respect to the Gaussian basis size, and we benchmark the performance of sc$GW$ on GPU architectures. 
In addition, in the absence of quasiparticle and non-self-consistent approximations, our work provides reference values of sc$GW$ band gaps for selected semiconductors and insulators that are compared to another finite-temperature sc$GW$ implementation~\cite{scGW_also_vertex_Andrey_2017}, where the numerical setup is entirely different. 

The paper will proceed as follows. 
In Sec.~\ref{sec:H_and_basis}, we introduce the electronic Hamiltonian in the context of Gaussian Bloch orbitals. 
Sec.~\ref{sec:scGW} discusses the self-consistent $GW$ equations as well as electronic thermodynamic properties, and Sec.~\ref{sec:implementation} describes details of our implementation, including the parallelization scheme employing GPUs. 
Lastly, Sec.~\ref{sec:results} compares our sc$GW$ data to other $GW$ implementations on a series of benchmark results. 
Our conclusions are presented in Sec.~\ref{sec:conclusion}. 

\section{Electronic Hamiltonian and finite basis sets\label{sec:H_and_basis}}
\subsection{Electronic Hamiltonian}

We solve a general electronic Hamiltonian $\hat{H} = \hat{H}_{0} + \hat{U}$, consisting of a one-electron part $\hat{H}_{0}$ and two-electron Coulomb interactions $\hat{U}$ (also known as electron repulsion integrals). 
In a translation invariant system, $\hat{H}$ in second quantization is
\begin{align}
&\hat{H} = \sum_{\bold{k}}\sum_{ij}\sum_{\sigma\sigma'}(H_{0})^{\bold{k}}_{i\sigma,j\sigma'}\hat{c}^{\bold{k}\dag}_{i\sigma}\hat{c}^{\bold{k}}_{j\sigma'} \\
&+ \frac{1}{2N_{k}}\sum_{ijkl}\sum_{\bold{k}_{i}\bold{k}_{j}\bold{k}_{k}\bold{k}_{l}}\sum_{\sigma\sigma'}U^{\bold{k}_{i}\bold{k}_{j} \bold{k}_{k}\bold{k}_{l}}_{\ i\ j\ \ k \ l}\hat{c}^{\bold{k}_{i}\dag}_{i\sigma}\hat{c}^{\bold{k}_{k}\dag}_{k\sigma'}\hat{c}^{\bold{k}_{l}}_{l\sigma'}\hat{c}^{\bold{k}+{j}}_{j\sigma} \nonumber,
\label{Eq:H}
\end{align}
where $N_{k}$ is the number of $\bold{k}$-points sampled in the Brillouin zone, and $\hat{c}^{\bold{k}\dag}_{i\sigma}$ ($\hat{c}^{\bold{k}}_{i\sigma}$) are the creation (annihilation) operators for electrons in the single-particle spin-orbital basis with crystal momentum $\bold{k}$, spin $\sigma$, and finite basis index $i$. 
The two-electron Coulomb interactions conserve crystal momentum, i.e. $\bold{k}_{i} + \bold{k}_{k} - \bold{k}_{j} - \bold{k}_{l} = \bold{G}$, where $\bold{G}$ is a reciprocal lattice vector. 
In general, $H_{0}$ exhibits a spin dependence with non-zero off-diagonal spin components. 
These components appear in the presence of spin-orbit coupling (SOC) and external magnetic fields, see Ref.~\onlinecite{scGW_X2C1e_Yeh}. 

We use bold symbols for matrices in the spin-orbital basis, bold italic symbols for tensors such as the two-electron Coulomb interactions $\boldsymbol{U}$, and regular italic symbols for matrix/tensor elements.

\subsection{Gaussian-type orbitals} 
We employ Bloch wave functions $g^{\bold{k}}_{i}(\bold{r})$ constructed from Gaussian-type orbitals (GTOs) $g^{\bold{R}}_{i}(\bold{r})$ as our finite basis set~\cite{CCSD_Bloch_GTO_2017,CRYSTAL_2018,PySCF_2020,CP2K_2020}. The Gaussian Bloch basis $g^{\bold{k}}_{i}(\bold{r})$ is expressed as 
\begin{align}
g^{\bold{k}}_{i}(\bold{r}) = \sum_{\bold{R}} g^{\bold{R}}_{i}(\bold{r})e^{i\bold{k}\cdot\bold{R}},
\end{align}
where $\bold{k}$ is a crystal momentum in the first Brillouin zone of the reciprocal space, and $g^{\bold{R}}_{i}(\bold{r})$ is the $i$-th Gaussian atomic orbital centered in unit cell $\bold{R}$~\cite{Boys_Gaussian_basis_1950}. The summation over $\bold{R}$ extends to the whole lattice. 

Gaussian Bloch waves for different crystal momenta are orthogonal. 
In orbital space, they define the overlap matrix 
\begin{align}
S^{\bold{k}}_{ij} = \int_{\Omega} d\bold{r} g^{\bold{k}*}_{i}(\bold{r})g^{\bold{k}}_{j}(\bold{r})
\label{Eq:scalar_S}
\end{align}
where $\Omega$ denotes the unit cell volume. 

In the non-relativistic case, the one-electron Hamiltonian $H_{0}$ and the two-electron Coulomb integrals $U$ are defined as 
\begin{align}
(&H_{0})^{\bold{k}}_{i\sigma,j\sigma'} = (H_{0})^{\bold{k}}_{ij}\delta_{\sigma\sigma'}\nonumber\\
&=\delta_{\sigma\sigma'}\int_{\Omega}d\bold{r}g^{\bold{k}*}_{i}(\bold{r})\Big[-\frac{1}{2}\nabla_{\bold{r}}^{2} + \sum_{\alpha}\frac{Z_{\alpha}}{|\bold{r}-\bold{r}_{\alpha}|}\Big]g^{\bold{k}}_{j}(\bold{r})
\end{align}
and 
\begin{align}
&U^{\bold{k}_{i}\bold{k}_{j} \bold{k}_{k}\bold{k}_{l}}_{\ i\ j\ \ k\ l} = \label{Eq:U}\\ 
&\int\int d\bold{r}d\bold{r}' g^{\bold{k}_{i}*}_{i}(\bold{r})g^{\bold{k}_{j}}_{j}(\bold{r})\frac{1}{|\bold{r}-\bold{r}'|}g^{\bold{k}_{k}*}_{k}(\bold{r}')g^{\bold{k}_{l}}_{l}(\bold{r}').  \nonumber
\end{align}
Ref.~\onlinecite{scGW_X2C1e_Yeh} derives the corresponding expression for a relativistic one-electron Hamiltonian in the exact two-component theory with one-electron approximation (X2C1e). 

\subsection{Decomposition of two-electron Coulomb interactions \label{subsec:decompose_Coulomb}}
The two-electron Coulomb interaction, Eq.~\ref{Eq:U}, has a memory requirement of $\mathcal{O}(N_{k}^{3}N_{orb}^{4})$ where $N_{orb}$ is the number of atomic orbitals in the unit cell. 
To reduce the size of this tensor, we use Coulomb potential decompositions. These decompositions can in general be expressed as 
\begin{align}
U^{\bold{k}_{i}\bold{k}_{j}\bold{k}_{k}\bold{k}_{l}}_{\ i\ j\ \ k\ l} = \sum_{Q}V^{\bold{k}_{i}\bold{k}_{j}}_{\ i\ j}(Q)V^{\bold{k}_{k}\bold{k}_{l}}_{\ k\ l}(Q), 
\label{Eq:U_decompose}
\end{align}
where $Q$ denotes an auxiliary decomposition index and $V^{\bold{k}_{i}\bold{k}_{j}}_{\ i\ j}(Q)$ is a tensor with two momenta, two orbital indices, and an auxiliary index. 
Decomposition procedures include the Cholesky decomposition~\cite{Cholesky_Boman2008} and the density fitting technique (also known as the resolution-of-identity (RI) approximation)~\cite{DF_Werner2003,DF_Ren2012,RSDF_HongZhou2021}. 
In the present work, we employ periodic Gaussian density fitting (GDF) with the overlap metric~\cite{RI_S_ref1_1993,RI_S_ref2_2016,RI_S_ref3_2017}. 
Given an additional set of auxiliary Gaussian orbitals $\chi^{\bold{q}}_{Q}(\bold{r})$ as the auxiliary basis, Eq.~\ref{Eq:U_decompose} is computed as 
\begin{align}
V^{\bold{k}_{i}\bold{k}_{j}}_{\ i \ j}(Q) = \sum_{PP'} (\bold{J}^{\bold{q}})^{1/2}_{QP'}(\bold{A}^{\bold{q}})^{-1}_{P'P}B^{\bold{k}_{i}\bold{k}_{j}}_{\ i\ j}(P). 
\label{Eq:VijQ}
\end{align}
where $\bold{q} = \bold{k}_{j}-\bold{k}_{i}$, and 
\begin{align}
A^{\bold{q}}_{P'P} &= \int_{\Omega} d\bold{r}\chi^{\bold{q}*}_{P}(\bold{r})\chi^{\bold{q}}_{P'}(\bold{r}), \label{Eq:2c1e_ovlp} \\ 
B^{\bold{k}_{i}\bold{k}_{j}}_{\ i\ j}(P) &= \int_{\Omega} d\bold{r} \chi^{\bold{q}*}_{P}(\bold{r}) g^{\bold{k}_{i}*}_{i}(\bold{r})g^{\bold{k}_{j}}_{j}(\bold{r}), \label{Eq:3c1e_ovlp} \\
J^{\bold{q}}_{PQ} &= \int \int d\bold{r}d\bold{r}' \frac{\chi^{\bold{q}*}_{P}(\bold{r})\chi^{\bold{q}}_{Q}(\bold{r}')}{|\bold{r}-\bold{r}'|}.
\label{Eq:DF_J}
\end{align}
 
 We choose the even-tempered basis (ETB) with the default progression parameter $\beta_{\mathrm{ETB}}=2.0$~\cite{ETB_2017} for $\chi^{\bold{q}}_{Q}(\bold{r})$.
 The typical size of the auxiliary basis $\{\chi^{\bold{q}}_{Q}(\bold{r})\}$ is roughly $3\sim10$ times size of the GTO basis $\{g^{\bold{k}}_{i}(\bold{r})\}$~\cite{GTO_review_Hill2012}.
Thus Eq.~\ref{Eq:VijQ} provides a much more compact representation of two-electron Coulomb interactions than Eq.~\ref{Eq:U_decompose}. 
In addition to lowering memory requirements, the decomposed three-index tensor reduces the scaling of sc$GW$, see Sec.~\ref{sec:scGW}. 

\section{Self-consistent finite-temperature $GW$ \label{sec:scGW}}
The central objects of finite-temperature perturbation theory are finite-temperature one-particle Green's functions $G^{\bold{k}}_{i\sigma,j\sigma}(\tau)$ and self-energies $\Sigma^{\bold{k}}_{i\sigma,j\sigma}(\tau)$. The Green's function $G^{\bold{k}}_{i\sigma,j\sigma}(\tau)$ is defined as 
\begin{align}
G^{\bold{k}}_{i\sigma,j\sigma}(\tau) &= -\frac{1}{Z}\text{Tr}\Big[ e^{-(\beta-\tau)(\hat{H}-\mu\hat{N})}\hat{c}^{\bold{k}}_{i\sigma}e^{-\tau(\hat{H}-\mu\hat{N})}\hat{c}^{\bold{k},\dag}_{j\sigma}\Big], \label{Eq:G}\\
Z &= \text{Tr}\Big[e^{-\beta(\hat{H}-\mu\hat{N})}\Big], \label{Eq:Z}
\end{align}
where $Z$ is the partition function, $\beta$ the inverse temperature, $\mu$ the chemical potential, $\hat{N}$ the particle-number operator, and $\tau\in [0,\beta]$ the imaginary time. 
Fourier transforms between imaginary-time and Matsubara frequency are defined as 
\begin{align}
G^{\bold{k}}_{i\sigma,j\sigma}(i\omega_{n}) &= \int^{\beta}_{0}d\tau G^{\bold{k}}_{i\sigma,j\sigma}(\tau) e^{i\omega_{n}\tau} \label{Eq:tau_to_iw}
\end{align}
and 
\begin{align}
G^{\bold{k}}_{i\sigma,j\sigma}(\tau) &= \frac{1}{\beta}\sum_{n}G^{\bold{k}}_{i\sigma,j\sigma}(i\omega_{n})e^{-i\omega_{n}\tau} \label{Eq:iw_to_tau}
\end{align}
where $\omega_{n} = (2n+1)\pi/\beta$, $n \in \mathbb{Z}$, are fermionic Matsubara frequencies. 
Given an interacting Green's function $\bold{G}^{\bold{k}}(\tau)$, the correlated density matrix is $\boldsymbol{\gamma}^{\bold{k}} = -\bold{G}^{\bold{k}}(\tau=\beta^{-})$ and the total number of electrons $N_{e}$ is determined as 
\begin{align}
N_{e} = \frac{1}{N_{k}}\sum_{\bold{k}}\mathrm{tr}[\boldsymbol{\gamma}^{\bold{k}}\bold{S}^{\bold{k}}], 
\end{align} 
where the trace implies a sum over the diagonals in the spin-orbital space. 

In the Matsubara frequency domain, the Dyson equation relating self-energies to Green's functions is 
\begin{align}
[\bold{G}^{\bold{k}}(i\omega_{n})]^{-1} &= (i\omega_{n}+\mu)\bold{S}^{\bold{k}} - \bold{H}^{\bold{k}}_{0} - \bold{\Sigma}^{\bold{k}}(i\omega_{n})[\boldsymbol{G}] \nonumber\\
&= [\bold{G}^{\bold{k}}_{0}(i\omega_{n})]^{-1} - \bold{\Sigma}^{\bold{k}}(i\omega_{n})[\boldsymbol{G}] \label{Eq:Dyson}
\end{align}
where $[\bold{G}^{\bold{k}}_{0}(i\omega_{n})]^{-1} = (i\omega_{n}+\mu)\bold{S}^{\bold{k}} - \bold{H}^{\bold{k}}_{0}$ is the non-interacting Green's function of the one-electron Hamiltonian $\bold{H}^{\bold{k}}_{0}$ and $\bold{\Sigma}^{\bold{k}}[\boldsymbol{G}]$ is the self-energy which is a functional of the full interacting Green's function $\bold{G}^{\bold{k}}(i\omega_{n})$. 
The inverse is defined as a matrix inversion in spin-orbital space for any given $\bold{k}$ and $i\omega_{n}$. 

Self-consistent $GW$ yields a particular approximation $(\bold{\Sigma}^{GW})^{\bold{k}}[\boldsymbol{G}]$ of the exact self-energy. 
Separating the self-energy into its static and dynamical part, 
\begin{align}
(\bold{\Sigma}^{GW})^{\bold{k}}[\boldsymbol{G}] (i\omega_n)= (\bold{\Sigma}^{GW}_{\infty})^{\bold{k}}[\boldsymbol{G}] + (\bold{\tilde{\Sigma}}^{GW})^{\bold{k}}[\boldsymbol{G}](i\omega_n), 
\label{eqn:2cGW_selfenergy}
\end{align}
$(\bold{\Sigma}^{GW}_{\infty})^{\bold{k}}$ is the static Hartree-Fock (HF) self-energy, and $(\bold{\tilde{\Sigma}}^{GW})^{\bold{k}}(i\omega_n)$ corresponds to the frequency-dependent $GW$ self-energy which is obtained via the summation of an infinite series of RPA-like `bubble' diagrams \cite{Hedin65}. 
Together with the Dyson equation (Eq.~\ref{Eq:Dyson}), the self-energy (Eq.~\ref{eqn:2cGW_selfenergy}) is solved as a functional of the interacting Green's function $\bold{G}^{\bold{k}}(i\omega_{n})$ iteratively until self-consistency between $\bold{G}^{\bold{k}}(i\omega_{n})$ and $(\bold{\Sigma}^{GW})^{\bold{k}}(i\omega_{n})$ is achieved. 

\subsection{Hartree-Fock self-energy}
The Hartree-Fock self-energy is static and can be further divided into a Hartree term ($\boldsymbol{J}$) and an exchange term ($\boldsymbol{K}$)
\begin{align}
(\Sigma^{GW}_{\infty})^{\bold{k}}_{i\sigma,j\sigma} = J^{\bold{k}}_{i\sigma,j\sigma} + K^{\bold{k}}_{i\sigma,j\sigma}. 
\label{Eq:HF_selfenergy}
\end{align}
Individually, each term can be expressed in terms of the GDF Coulomb tensor $V^{\bold{k}\bold{k}'}_{\ i j}(Q)$ and the density matrix $\gamma^{\bold{k}'}_{b\sigma,a\sigma'}$ as 
\begin{align}
J^{\bold{k}}_{i\sigma,j\sigma} &= \frac{1}{N_{k}}\sum_{\bold{k}'}\sum_{\sigma_{1}}\sum_{ab} \gamma^{\bold{k}'}_{a\sigma_{1},b\sigma_{1}}U^{\bold{k}\bold{k}\bold{k}'\bold{k}'}_{i \ j \  b \ a} \\
&= \frac{1}{N_{k}}\sum_{\bold{k}'}\sum_{\sigma_{1}}\sum_{ab}\sum_{Q}V^{\bold{k}\bold{k}}_{\ ij}(Q)\gamma^{\bold{k}'}_{a\sigma_{1},b\sigma_{1}}V^{\bold{k}'\bold{k}'}_{\ b\ a}(Q)
\label{Eq:HF_J}
\end{align}
and 
\begin{align}
K&^{\bold{k}}_{i\sigma,j\sigma'} = -\frac{1}{N_{k}}\sum_{\bold{k}'}\sum_{ab} \gamma^{\bold{k}'}_{a\sigma,b\sigma'}U^{\bold{k}\bold{k}'\bold{k}'\bold{k}}_{i \ a \ b \ j}\\
&= -\frac{1}{N_{k}}\sum_{\bold{k}'}\sum_{ab}\sum_{Q}V^{\bold{k}\bold{k}'}_{\ i\ a}(Q)\gamma^{\bold{k}'}_{a\sigma,b\sigma'}V^{\bold{k}'\bold{k}}_{\ b\ j}(Q).
\label{Eq:HF_K}
\end{align}

\subsection{Dynamical part of the $GW$ self-energy}

\begin{figure}[tb!]
\begin{center}
\includegraphics[width=0.45\textwidth]{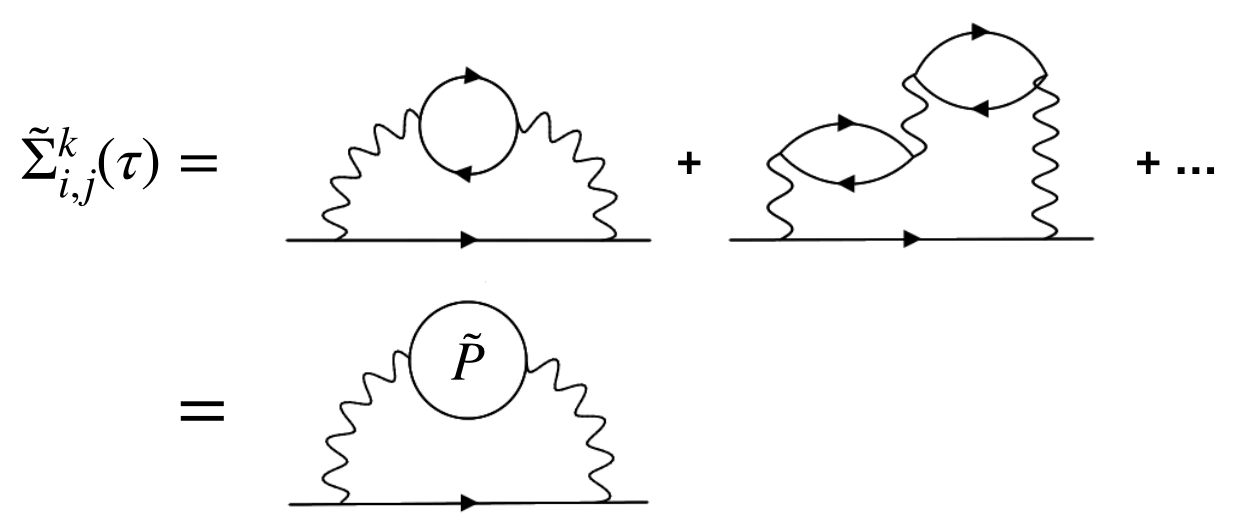}
\caption{Self-energy diagrams for the dynamical part of the $GW$ self-energy. Lines with arrows denote interacting Green's function $G^{\bold{k}}_{i\sigma,j\sigma}$, wiggly lines denote the interaction $U^{\bold{k},\bold{k}-\bold{q}, \bold{k}',\bold{k}'+\bold{q}}_{\ i,\ \ j,\ \ k,\ \ \ l}$. $\boldsymbol{\tilde{P}}$ represents the sum of all ``bubble" diagrams.}\label{Fig:GW_selfenergy}
\end{center}
\end{figure}
\begin{figure}[tb!]
\begin{center}
\includegraphics[width=0.48\textwidth]{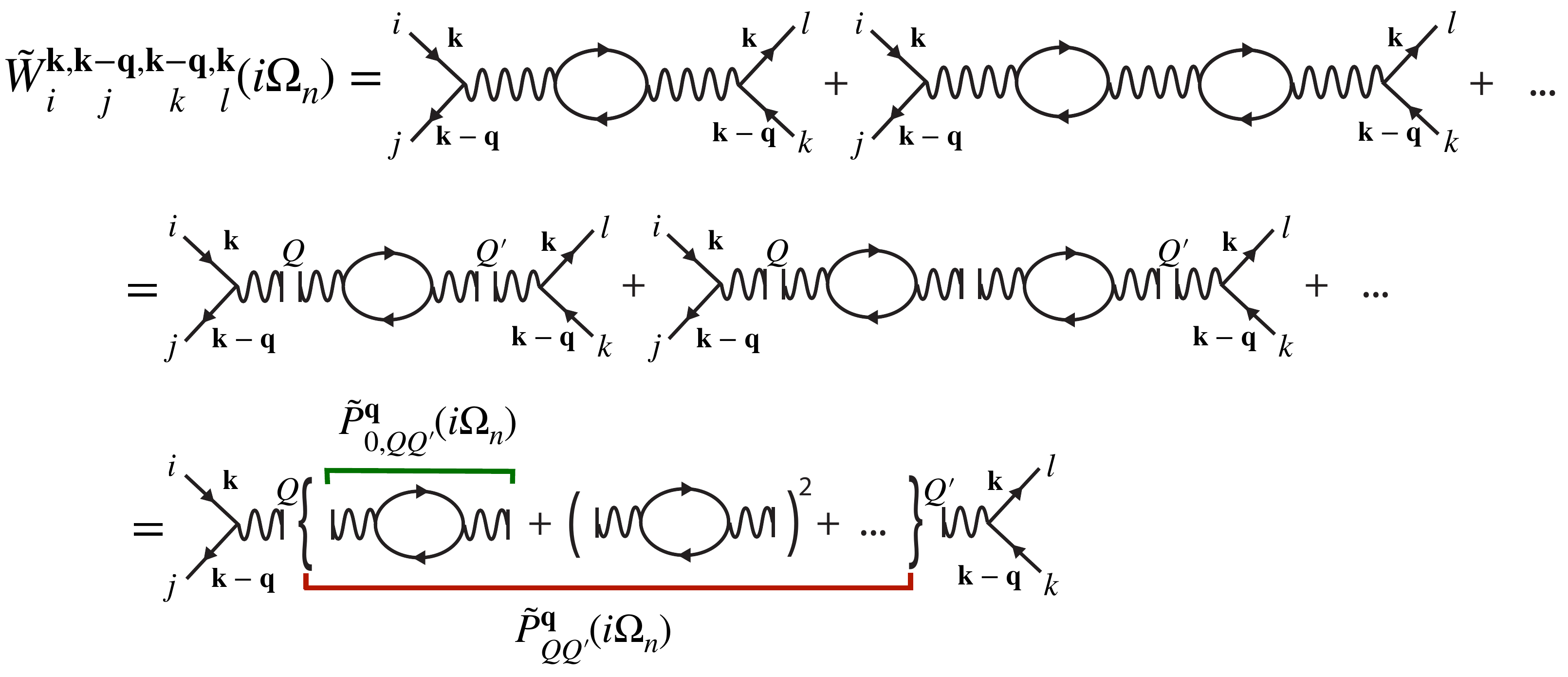}
\caption{Diagrammatic expression of $\boldsymbol{\tilde{W}}$.}\label{Fig:tilde_W}
\end{center}
\end{figure}

In $GW$, the dynamical part of the self-energy is approximated as the sum of an infinite series of RPA-like `bubble' diagrams~\cite{Hedin65} as shown in Fig~\ref{Fig:GW_selfenergy}. 
On the imaginary-time axis, $(\tilde{\bold{\Sigma}}^{GW})^{\bold{k}}(\tau)$ reads 
\begin{align}
(\tilde{\Sigma}^{GW})^{\bold{k}}_{i\sigma,j\sigma}(\tau) &= -\frac{1}{N_{k}}\sum_{\bold{q}}\sum_{ab} G^{\bold{k-q}}_{a\sigma,b\sigma}(\tau)\tilde{W}^{\bold{k},\bold{k-q},\bold{k-q},\bold{k}}_{\ i\ \ a \ \ \ \ b \ \ \ j}(\tau)
\label{Eq:tilde_sigma}
\end{align}
where $\boldsymbol{\tilde{W}}$ is the effective screened interaction tensor, defined as the difference between the full dynamically screened interaction $\boldsymbol{W}$ and the bare interaction $\boldsymbol{U}$, i.e. $\boldsymbol{\tilde{W}} = \boldsymbol{W} - \boldsymbol{U}$. 
In the $GW$ approximation, the screened interaction $\boldsymbol{W}$ is expressed as~\cite{Hedin65} 
\begin{align}
&W^{\bold{k}_{1}\bold{k}_{2}\bold{k}_{3}\bold{k}_{4}}_{\ i\ \ j \ k\  \ l}(i\Omega_{n}) = U^{\bold{k}_{1}\bold{k}_{2}\bold{k}_{3}\bold{k}_{4}}_{\ i\ \ j \ k\  \ l}\nonumber\\
&+\frac{1}{N_{k}}\sum_{\bold{k}_{5}\bold{k}_{6}\bold{k}_{7}\bold{k}_{8}}\sum_{abcd}U^{\bold{k}_{1}\bold{k}_{2}\bold{k}_{5}\bold{k}_{6}}_{\ i\ \ j \ a\  \ b} \mathit{\Pi}^{\bold{k}_{5}\bold{k}_{6}\bold{k}_{7}\bold{k}_{8}}_{\ a\ b \ c\  \ d}(i\Omega_{n})W^{\bold{k}_{7}\bold{k}_{8}\bold{k}_{3}\bold{k}_{4}}_{\ c\ d \ \ k\  \ l}(i\Omega_{n})
\label{Eq:W}
\end{align}
where $\boldsymbol{\mathit{\Pi}}$ is the non-interacting polarization function 
\begin{align}
\mathit{\Pi^{\bold{k}_{1}\bold{k}_{2}\bold{k}_{3}\bold{k}_{4}}_{\ a\ b \ c\  \ d}(\tau)} = \sum_{\sigma}G^{\bold{k}_{1}}_{d\sigma,a\sigma}(\tau)G^{\bold{k}_{2}}_{b\sigma ,c\sigma}(-\tau)\delta_{\bold{k}_{1}\bold{k}_{4}}\delta_{\bold{k}_{2}\bold{k}_{3}}. 
\label{Eq:P0_tau}
\end{align}
Due to the size of the interaction tensor $\boldsymbol{U}$, solving Eq.~\ref{Eq:W} and~\ref{Eq:P0_tau} directly is not practical in large simulations. 
As illustrated in Fig.~\ref{Fig:tilde_W}, the decomposition of Coulomb integrals allows us to express $\boldsymbol{\tilde{W}}$ as 
\begin{align}
&\tilde{W}^{\bold{k},\bold{k-q},\bold{k-q},\bold{k}}_{\ i\ \ j \ \ \ \ k \ \ \ l}(i\Omega_{n}) = \sum_{Q,Q'}V^{\bold{k},\bold{k-q}}_{\ i\ \ j}(Q)\times\nonumber\\
&\Big\{ \tilde{P}^{\bold{q}}_{0,QQ'}(i\Omega_{n}) + \big[\tilde{P}^{\bold{q}}_{0}(i\Omega_{n})\big]^{2}_{QQ'}+...\Big\}V^{\bold{k-q},\bold{k}}_{\ \ k \ \ \ l}(Q')
\label{Eq:tilde_W_DF}\\
&=\sum_{Q,Q'}V^{\bold{k},\bold{k-q}}_{\ i\ \ j}(Q)\tilde{P}^{\bold{q}}_{QQ'}(i\Omega_{n})V^{\bold{k-q},\bold{k}}_{\ \ k \ \ \ l}(Q')
\label{Eq:tilde_W_DF2}
\end{align}
where $\Omega_{n} = 2n\pi/\beta$ ($n \in \mathbb{Z}$) are bosonic Matsuabara frequencies. 
The non-interacting auxiliary function $\tilde{\bold{P}}^{\bold{q}}_{0}(i\Omega_{n})$ is defined as 
\begin{subequations}
\label{Eqs:P0}
\begin{align}
\tilde{P}^{\bold{q}}_{0,QQ'}(i\Omega_{n}&) = \int_{0}^{\beta}d\tau \tilde{P}^{\bold{q}}_{0,QQ'}(\tau)e^{i\Omega_{n}\tau}, \label{Eq:P0t_to_P0w}\\
\tilde{P}^{\bold{q}}_{0,QQ'}(\tau)& = \frac{-1}{N_{k}}\sum_{\bold{k}}\sum_{\sigma\sigma'}\sum_{abcd}V^{\bold{k},\bold{k+q}}_{\ d \ a}(Q)\nonumber\\
&\times G^{\bold{k}}_{c\sigma',d\sigma}(-\tau)G^{\bold{k+q}}_{a\sigma \ b\sigma'}(\tau)V^{\bold{k+q},\bold{k}}_{\ \ b\  \ \ c}(Q'),\label{Eq:tilde_P0_tau}
\end{align}
\end{subequations}
and the renormalized auxiliary function $\tilde{\bold{P}}^{\bold{q}}(i\Omega_{n})$ is computed using the geometric series 
\begin{align}
\tilde{\bold{P}}^{\bold{q}}(i\Omega_{n}) = \sum_{m=1}^{\infty}[\tilde{\bold{P}}^{\bold{q}}_{0}(i\Omega_{n})]^{m} 
= [\bold{I} - \tilde{\bold{P}}^{\bold{q}}_{0}(i\Omega_{n})]^{-1}\tilde{\bold{P}}^{\bold{q}}_{0}(i\Omega_{n})
\label{Eq:tilde_P_omega}
\end{align}
where the inverse denotes a matrix inversion in the auxiliary orbital space and $\bold{I}$ is a unitary matrix. 
Transforming $\tilde{\bold{P}}^{\bold{q}}(i\Omega_{n})$ from the Matsubara frequency to the imaginary-time domain 
\begin{align}
&\tilde{P}^{\bold{q}}_{QQ}(\tau) = \frac{1}{\beta}\sum_{n}\tilde{P}^{\bold{q}}_{QQ'}(i\Omega_{n})e^{-i\Omega_{n}\tau},
\label{Eq:Piw_to_Pt}
\end{align}
and then inserting it into Eq.~\ref{Eq:tilde_sigma}, we arrive at 
\begin{align}
(&\tilde{\Sigma}^{GW})^{\bold{k}}_{i\sigma,j\sigma}(\tau) = \nonumber \\
&\frac{-1}{N_{k}}\sum_{\bold{q}}\sum_{ab}\sum_{QQ'}G^{\bold{k-q}}_{a\sigma,b\sigma}(\tau)V^{\bold{k},\bold{k-q}}_{\ i \ \ a}(Q)\tilde{P}^{\bold{q}}_{QQ'}(\tau)V^{\bold{k-q},\bold{k}}_{\ \ b \ \ \ j}(Q').
\label{Eq:tilde_sigma2}
\end{align}

The dynamical self-energy, expressed in such a way, is then evaluated directly on the imaginary-time axis. 
Self-consistent iterations on the imaginary axis are straightforward and stable. 
Further approximations to the placement of poles, ``quasi-particle'' approximations, or approximations to the off-diagonal self-energy structure are not needed. 
However, the evaluation of real-frequency spectra and band gaps requires an analytical continuation to real frequencies. 
Recently developed complex analysis techniques~\cite{Nevanlinna_Jiani_2021,Caratheodory_Jiani_2021} can be employed to perform this step accurately. 

\subsection{Thermodynamic properties \label{subsec:thermodynamics}}

\begin{figure}[tb!]
\begin{center}
\includegraphics[width=0.45\textwidth]{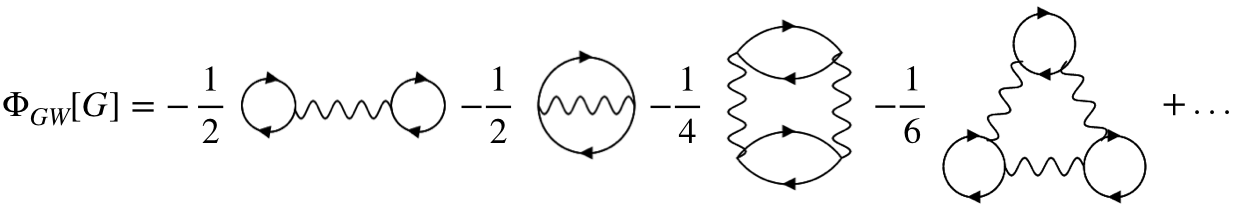}
\caption{Diagrammatic expansion of $\Phi$-function in the $GW$ approximation.}\label{Fig:Phi_GW}
\end{center}
\end{figure}

For a conserving approximation, the grand potential $\Omega$ is defined in terms of a Green's function, a self-energy, and the corresponding $\Phi$ functional as~\cite{LW_functional_Luttinger1960}
\begin{align}
\Omega[\boldsymbol{G}] = \Phi[\boldsymbol{G}] - \mathrm{Tr}\{ \mathrm{ln}[-\boldsymbol{G}^{-1}\} - \mathrm{Tr}\big\{\boldsymbol{\Sigma}[\boldsymbol{G}]\boldsymbol{G}\big\}, 
\label{Eq:GP}
\end{align}
where the symbol $\mathrm{Tr}\{\}$ includes summations over crystal momentum ($\frac{1}{N_{k}}\sum_{\bold{k}}$), Matsubara frequency ($\frac{1}{\beta}\sum_{n}$), and spin-orbital index $(i, \sigma)$. 
The Luttinger-Ward functional $\Phi[\boldsymbol{G}]$ is a functional of $\boldsymbol{G}$ expressed as 
\begin{align}
\Phi[\boldsymbol{G}] = \sum_{m=1}^{\infty} \frac{1}{2m}\mathrm{Tr} \big\{ \boldsymbol{\mathit{\Sigma}}^{(m)}[\boldsymbol{G}] \boldsymbol{G} \big\}
\end{align}
where $\boldsymbol{\mathit{\Sigma}}^{(m)}[\boldsymbol{G}]$ is the $m$-th order skeleton diagram of the self-energy $\boldsymbol{\mathit{\Sigma}}$. 
Within the $GW$ approximation, the $\Phi$ functional, as shown in Fig.~\ref{Fig:Phi_GW}, is expressed as  
\begin{align}
\Phi_{GW}[G] &= -\frac{1}{2}\mathrm{Tr} \{\boldsymbol{\mathit{\Sigma}}^{GW}_{\infty}\boldsymbol{\mathit{\gamma}}\} - \frac{1}{4}\mathrm{Tr}\{(\boldsymbol{\mathit{U\Pi}})^{2}\} + ...\label{Eq:Phi_GW}\\
&= \Phi^{GW}_{\infty} + \tilde{\Phi}_{GW} \label{Eq:Phi_GW}
\end{align}
where $\boldsymbol{U}$ are the bare Coulomb integrals (Eq.~\ref{Eq:U}) and $\boldsymbol{\mathit{\Pi}}$ is the non-interacting polarization function (Eq.~\ref{Eq:P0_tau}). 
We define the first term in Eq.~\ref{Eq:Phi_GW} as the contribution of the static $GW$ self-energy $\Phi^{GW}_{\infty}$ and attribute the rest coming from the dynamical $GW$ self-energy diagrams as $\tilde{\Phi}_{GW}$. 

Using the decomposed Coulomb integrals (Eq.~\ref{Eq:U_decompose}), the first term in $\tilde{\Phi}_{GW}$ is rewritten as 
\begin{align}
\frac{1}{4}&\mathrm{Tr} \{(\boldsymbol{\mathit{U\Pi}})^{2}\}  \\
&=\frac{1}{4N_{k}}\sum_{\bold{q}}\frac{1}{\beta}\sum_{n}\sum_{QQ'}\tilde{P}^{\bold{q}}_{0,QQ'}(i\Omega_{n})\tilde{P}^{\bold{q}}_{0,Q'Q}(i\Omega_{n}) \nonumber \\
&=\frac{1}{4N_{k}}\sum_{\bold{q}}\frac{1}{\beta}\sum_{n}\mathrm{tr}\big\{\tilde{\bold{P}}^{\bold{q}}_{0}(i\Omega_{n})^{2}\big\} \nonumber \\
&=\frac{1}{4}\mathrm{Tr}\{(\boldsymbol{\tilde{P}}_{0})^{2}\} \nonumber
\end{align}
where $\boldsymbol{\tilde{P}}_{0}$ is the non-interacting auxiliary function defined in Eqs.~\ref{Eqs:P0} and $\mathrm{tr}\{\}$ represents the trace of a matrix in the auxiliary Gaussian orbital space. 
Similarly, $\tilde{\Phi}_{GW}$ can be rewritten in terms of the non-interacting auxiliary function $\boldsymbol{\tilde{P}}_{0}$ as 
\begin{align}
&\tilde{\Phi}_{GW} = -\frac{1}{2} \Big( [\mathrm{Tr}\{\boldsymbol{\tilde{P}}_{0}\} + \frac{1}{2}\mathrm{Tr}\{(\boldsymbol{\tilde{P}}_{0})^{2}\} + ...] - \mathrm{Tr}\{\boldsymbol{\tilde{P}}_{0}\}\Big) \nonumber\\
&=\frac{1}{2N_{k}}\sum_{\bold{q}}\frac{1}{\beta}\sum_{n}\mathrm{tr}\{\mathrm{ln}[\bold{I} - \tilde{\bold{P}}^{\bold{q}}_{0}(i\Omega_{n})] + \tilde{\bold{P}}^{\bold{q}}_{0}(i\Omega_{n})\}.\label{Eq:tilde_Phi_GW}
\end{align}

Inserting Eq.~\ref{Eq:Phi_GW} and~\ref{Eq:tilde_Phi_GW} into Eq.~\ref{Eq:GP}, the $GW$ grand potential is defined. Other thermodynamic quantities such as entropies, specific heats, total energies and free energies are then evaluated using standard thermodynamic expressions~\cite{LW_functional_Luttinger1960,GM_Energy_Holm_2000,Fetter_Walecka_MBTheory_2003,Variational_energy_functionals_van_Leeuwen_2006,GF2_Sergei_2019}. 

In $\Phi$-derivable self-consistent methods such as sc$GW$ and the self-consistent second-order Green's function perturbation theory (GF2)~\cite{Alicia_GF2_2016,GF2_Sergei_2019}, 
 thermodynamic quantities are independent of the integration path and the choice of method~\cite{Baym_Kadanoff_1961,Baym_1962}.
For example, different approaches for obtaining the total energy from the single-particle Green's function, such as using the Galitskii-Migdal formula or thermodynamic integration, all lead to the same result when the self-energy is $\Phi$-derivable~\cite{Baym_Kadanoff_1961,Baym_1962,Variational_energy_functionals_van_Leeuwen_2006}.  
$\Phi$-derivable approximations also guarantee conservation of particle number, momentum, and energy~\cite{Baym_1962}.

\subsection{Treatment of the integrable divergence \label{subsec:head_corr}}

The two-electron Coulomb integral, Eq.~\ref{Eq:U_decompose}, has a singularity at $\bold{q}\rightarrow\bold{0}$ due to the long-range contribution of the Coulomb kernel $\sim1/|\bold{q+G}|^{2}$ at $\bold{G}=\bold{0}$. 
In calculations of the GDF Coulomb tensors $V^{\bold{k}_{1}\bold{k}_{2}}_{\ i\ \ j}(Q)$ (Eq.~\ref{Eq:VijQ}), the $\bold{G}=\bold{0}$ contribution is manually excluded  at $\bold{q=0}$ to regularize the divergence. 

In the Coulomb potential $\boldsymbol{J}$ (Eq.~\ref{Eq:HF_J}), the singularity can be safely excluded since the divergence is cancelled exactly by the electron-nucleus Coulomb potential counterpart. 
When an infinite number of $\bold{k}$-points is sampled in the first Brillouin zone, the singularities are in fact integrable in both the HF exchange potential (Eq.~\ref{Eq:HF_K}) and the dynamical $GW$ self-energy (Eq.~\ref{Eq:tilde_sigma}), resulting in finite contributions~\cite{Exchange_correction_auxiliary_function_Gygi_1986,ERI_correction_2005,ERI_correction_2009,GW_BZ_integration_around_gamma_Shishkin_2006,GW_offset_gamma_Kotani2007,GW_analytic_integration_Huser_2013,G0W0_solids_CP2K_Wilhelm_2017}. 
However, in practice, for any finite size $\bold{k}$-mesh, both Eq.~\ref{Eq:HF_K} and~\ref{Eq:tilde_sigma} exhibit a divergence. 
This can be understood by rewriting the effective screened interaction $\tilde{W}$ in Eq.~\ref{Eq:tilde_sigma} in terms of the plane-wave basis ($\bold{G}$), 
\begin{align}
\tilde{W}^{\bold{k},\bold{k-q},\bold{k-q},\bold{k}}_{\ i\ \ a \ \ \ \ b \ \ \ j}&(\tau) = \frac{1}{\Omega}\sum_{\bold{G}\bold{G}'} \rho^{\bold{k-q}\bold{k}*}_{\ \ a \ \ i}(\bold{G})\frac{\sqrt{4\pi}}{|\bold{q}+\bold{G}|}\label{Eq:W_GG'}\\
&\times(\epsilon^{\bold{q},-1}_{\bold{G}\bold{G}'}(\tau) - \delta_{\bold{G}\bold{G}'})\frac{\sqrt{4\pi}}{|\bold{q}+\bold{G}'|}\rho^{\bold{k-q}\bold{k}}_{\ \ b \ \ j}(\bold{G}'), \nonumber
\end{align}
where $\epsilon^{\bold{q}}_{\bold{G}\bold{G}'}(\tau)$ is the effective dielectric function in the plane-wave basis, and $\rho^{\bold{k}_{1}\bold{k}_{2}}_{\ i \ j}(\bold{G})$ is the Fourier transform of a GTO pair density function $\rho^{\bold{k}_{1}\bold{k}_{2}}_{\ i \ j}(\bold{r}) = g^{\bold{k}_{1}*}_{i}(\bold{r})g^{\bold{k}_{2}}_{j}(\bold{r})$, 
\begin{align}
\rho^{\bold{k}_{1}\bold{k}_{2}}_{\ i \ j}(\bold{G}) = \int_{\Omega}d\bold{r} \rho^{\bold{k}_{1}\bold{k}_{2}}_{\ i \ j}(\bold{r}) e^{-i(\bold{k}_{2}-\bold{k}_{1}+\bold{G})\bold{r}}. 
\end{align}
Inserting Eq.~\ref{Eq:W_GG'} into Eq.~\ref{Eq:tilde_sigma}, we obtain 
\begin{align}
(&\tilde{\Sigma}^{GW})^{\bold{k}}_{i\sigma,j\sigma}(\tau)  = \frac{-1}{N_{k}\Omega}\sum_{\bold{q}}\sum_{\bold{G}\bold{G}'}\sum_{ab} G^{\bold{k-q}}_{a\sigma,b\sigma}(\tau)\times \\ 
&\rho^{\bold{k-q}\bold{k}*}_{\ \ a \ \ i}(\bold{G})\frac{\sqrt{4\pi}}{|\bold{q}+\bold{G}|}(\epsilon^{\bold{q},-1}_{\bold{G}\bold{G}'}(\tau) - \delta_{\bold{G}\bold{G}'})\frac{\sqrt{4\pi}}{|\bold{q}+\bold{G}'|}\rho^{\bold{k-q}\bold{k}}_{\ \ b \ \ j}(\bold{G}'). \nonumber
\end{align}
At $\bold{q}\rightarrow\bold{0}$, the screened interaction in the plane-wave basis diverges when $\bold{G}=\bold{0}$ or $\bold{G}'=\bold{0}$.  
A similar divergence in the HF exchange potential can be understood by replacing $(\epsilon^{\bold{q},-1}_{\bold{G}\bold{G}'}(\tau) - \delta_{\bold{G}\bold{G}'})$ with $\delta_{\bold{G}\bold{G}'}$. 
The explicit exclusion of the $\bold{G=0}$ contribution in the two-electron Coulomb integrals avoids these divergences. 
At the same time, it will result in a slow convergence with respect to the number $\bold{k}$-points sampled, which impedes a rapid convergence to the thermodynamic limit (TDL) in practical calculations. 
A finite-size correction is therefore necessary to accelerate convergence to the TDL. 

Several strategies of correcting these finite-size effects have been proposed~\cite{Exchange_correction_auxiliary_function_Gygi_1986,ERI_correction_2005,ERI_correction_2009,GW_BZ_integration_around_gamma_Shishkin_2006,GW_offset_gamma_Kotani2007,GW_analytic_integration_Huser_2013,G0W0_solids_CP2K_Wilhelm_2017}. 
We follow the procedure of Gygi and Baldereschi~\cite{Exchange_correction_auxiliary_function_Gygi_1986} in which an auxiliary function is subtracted and added back on the right-hand side of Eq.~\ref{Eq:HF_K}, and~\ref{Eq:tilde_sigma}. 
The singularity is first removed by subtracting an auxiliary function that exhibits the same divergence $\sim 1/\bold{q}^{2}$ as $\bold{q}\rightarrow\bold{0}$. 
Therefore, the resulting smooth integrand can be evaluated accurately by a summation over  a finite number of $\bold{k}$-points, and the singularity is transferred to the added term expressed by the auxiliary function. 
The key point is that this added term can be analytically integrated. 
In principle, the choice of auxiliary function is arbitrary since convergence will be achieved upon increasing the number of $\bold{k}$-points, irrespective of the correction. 
However, a proper choice of auxiliary function will accelerate the convergence with respect to the number of $\bold{k}$-points. 

In the present work, the auxiliary function used to correct the HF exchange potential in Ref.~\onlinecite{ERI_correction_2009} is adopted for both Eq.~\ref{Eq:HF_K} and~\ref{Eq:tilde_sigma}. 
For the dynamical $GW$ self-energy, only the leading-order correction at $\bold{G}=\bold{G}'=\bold{0}$ is included, which is the so-called head correction 
\begin{align}
(\Delta&^{GW})^{\bold{k}}_{i\sigma,j\sigma'}(\tau) = -\chi \sum_{ab} G^{\bold{k-q}}_{a\sigma,b\sigma'}(\tau)\rho^{\bold{k-q}\bold{k}*}_{\ \ a \ \ i}(\bold{G})\nonumber\\
&\times[\epsilon^{\bold{q},-1}_{\bold{G}\bold{G}'}(-\tau) - \delta_{\bold{G}\bold{G}'}]\rho^{\bold{k-q}\bold{k}}_{\ b \ j}(\bold{G}')\Big|_{\bold{q}=\bold{G}=\bold{G}'=\bold{0}}\\
&= -\chi [\epsilon^{\bold{0},-1}_{\bold{00}}(-\tau) - 1] \sum_{ab}S^{\bold{k}}_{ia}G^{\bold{k}}_{a\sigma,b\sigma'}(\tau)S^{\bold{k}}_{bj}, 
\label{Eq:GW_head_corr}
\end{align}
where $\chi$ is the supercell Madelung constant~\cite{ERI_correction_2009}. 
The head correction is dynamical and requires the knowledge of the dielectric constant in the long-wavelength limit. 
However, a direct evaluation of $\epsilon^{\bold{q=0}}_{\bold{G=0,G'=0}}$ is not possible due to the singularity of the bare Coulomb interaction~\cite{scGW_also_vertex_Andrey_2017,scGW_VASP_2018}. 
Instead, we fit $\epsilon^{\bold{q}}_{\bold{G=0,G'=0}}$ using a least-square fit with a finite number of $\bold{q}$-points around the $\Gamma$-point, and then extrapolate to $\bold{q=0}$.  
Lastly, a static finite-size correction for the HF exchange potential reads 
\begin{align}
(\Delta^{\mathrm{HF}})^{\bold{k}}_{i\sigma,j\sigma'} = -\chi \sum_{ab}S^{\bold{k}}_{ia}\gamma^{\bold{k}}_{a\sigma,b\sigma'}S^{\bold{k}}_{bj}. 
\label{Eq:K_corr}
\end{align}
Additional details of the derivation of Eq.~\ref{Eq:GW_head_corr} and~\ref{Eq:K_corr} are shown in Appendix~\ref{appendix:div_treatment}. 

\section{Implementation details \label{sec:implementation}}
\subsection{sc$GW$ workflow \label{subsec:GW_workflow}}
The unit cell and crystal structure, the GTO basis set, the auxiliary basis set, the temperature, and the $\bold{k}$-mesh fully define the electronic structure problem and allow to precompute the one-electron Hamiltonian $(H_{0})^{\bold{k}}_{i\sigma,j\sigma'}$, overlap matrices $S^{\bold{k}}_{ij}$, and density-fitted Coulomb interactions $V^{\bold{k}\bold{k}'}_{\ i j}(Q)$. 
In the present work, all of the tensors $(H_{0})^{\bold{k}}_{i\sigma,j\sigma'}$, $S^{\bold{k}}_{ij}$, and $V^{\bold{k}\bold{k}'}_{\ i j}(Q)$ are pre-computed using the \texttt{PySCF} package~\cite{PySCF_2020} and stored on disk. 

\begin{figure}[tb!]
\begin{center}
\includegraphics[width=0.40\textwidth]{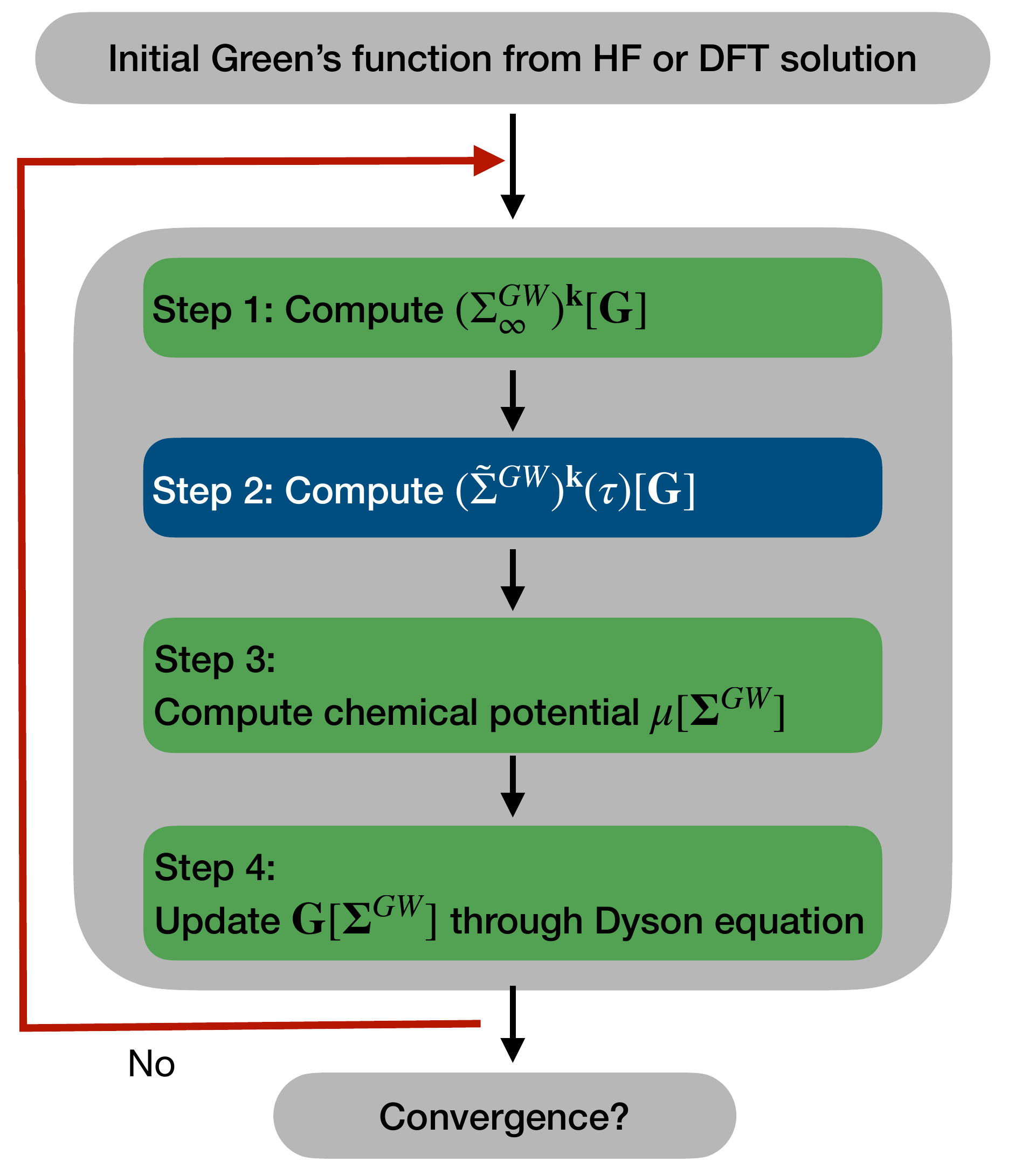}
\caption{Workflow of sc$GW$. Green boxes represent code segments that can be accelerated using MPI while the blue box contains computationally intensive parts implemented in a CPU-GPU hybrid architecture. }\label{Fig:scGW_workflow}
\end{center}
\end{figure}
Fig.~\ref{Fig:scGW_workflow} shows the workflow of our sc$GW$ implementation. 
Step 1 and 2 are functionals of $\bold{G}$ while step 3 and 4 are functionals of $\bold{\Sigma}^{GW}$. 
Starting from an initial guess for the Green's function (obtained, for instance, from Hartree-Fock or DFT),  the static Hartree-Fock self-energy $(\bold{\Sigma}^{GW}_{\infty})^{\bold{k}}$ is computed with Eq.~\ref{Eq:HF_J},~\ref{Eq:HF_K}. This Green's function is also used to compute the dynamical part of the $GW$ self-energy $(\tilde{\bold{\Sigma}}^{GW})^{\bold{k}}(\tau)$  through Eq.~\ref{Eq:tilde_sigma2}. The dynamical self-energy is then Fourier transformed to Matsubara space.
Next, using the newly computed $(\bold{\Sigma}^{GW}_{\infty})^{\bold{k}}$ and $(\tilde{\bold{\Sigma}}^{GW})^{\bold{k}}(i\omega_n)$, the chemical potential is adjusted in the presence of the new $GW$ self-energy such that the total number of electrons corresponds to a charge-neutral system. (We employ a threshold value of $10^{-9}$.)  
Lastly, the Green's function is updated via the Dyson equation and serves as an input to the next iteration. 

This self-consistent iteration is repeated until convergence in Eq.~\ref{Eq:Dyson} and~\ref{eqn:2cGW_selfenergy} is reached. 
We validate convergence in the Green's function and the self-energy as well as in the energy and in the chemical potential. 
The additional computational overhead compared to the one-shot $G_0W_0$ variant is proportional to the number of iterations required to reach a convergence. 
Typically 10$\sim$100 iterations are needed. 

\subsection{Convergence acceleration\label{subsec:iterative_solution}}
The number of iterations to reach a self-consistent solution directly affects the efficiency of a sc$GW$ implementation. 
Two aspects will affect the convergence behavior, the initial guess of the Green's function and the iterative solver. 

A ``good'' initial starting point will result in a fast and stable convergence. 
Besides the initial guess based on a bare non-interacting Hamiltonian $(H_{0})^{\bold{k}}_{i\sigma,j\sigma'}$, starting points such as the effective Hartree-Fock or DFT one-electron Green's functions are commonly used. 
For systems where multiple meta-stable states are present, an inadequate starting point may guide sc$GW$ to converge to a meta-stable state~\cite{multi_solution_GW_GF2_Pavel_2021}. 

Iterative numerical methods that facilitate and stabilize the convergence of self-consistent equations have been an active field of research. 
Common iterative solvers such as the direct inversion in the iterative subspace (DIIS)~\cite{Anderson_acceleration_1965,DIIS_Pulay_1980,DIIS_Pulay_1982,Anderson_acceleration_Walker2011} and the Newton method have been widely applied in DFT and other electronic structure methods for convergence of frequency-independent one-particle quantities~\cite{Anderson_acceleration_1965,DIIS_Pulay_1980,DIIS_Pulay_1982,DIIS_evGW,QSGW_molecules_Forster_2021}. 
In contrast, applications to Green's-function-based methods are so far limited. 
In the present work, we adopt a DIIS algorithm customized for finite-temperature solution of the Dyson equation~\cite{DIIS_Pavel_2022}. 

\subsection{Representation on the imaginary axes\label{subsec:IR}}
The representation of dynamical quantities critically affect the computational cost and the memory requirements. 
The compact representation of the frequency dependence of one-body and two-body quantities is an active field of research~\cite{Legendre_Boehnke_2011,IR_Hiroshi_2017,Chebyshev_Gull_2018,Legendre_Dong2020,Minimax_Kaltak_2020,sparse_sampling_Jia_2020,DLR_Kaye_2021}. 
In the present work, all dynamical quantities, including fermionic and bosonic functions, are expanded into the intermediate representation (IR)~\cite{IR_Hiroshi_2017}, generated using the \texttt{IRBASIS} open-source software package~\cite{irbasis_CPC_Chikano2019}, with sparse sampling on both the imaginary-time and Matsubara frequency axes~\cite{sparse_sampling_Jia_2020}. 
The intermediate representation is controlled by a dimensionless parameter $\lambda$ that should be chosen to be larger than $\beta\tilde{\omega}$ where $\beta$ is the inverse temperature and $\tilde{\omega}$ is the bandwidth of the system. 
The typical number of imaginary-time coefficients retained is 10 $\sim$ 200.

\subsection{Spectral function $A(\omega)$\label{subsec:spectral}}
Once the sc$GW$ single-particle Green's function has been computed, the $\bold{k}$-resolved spectral function $A^{\bold{k}}(\omega)$ can be extracted by inverting the relation 
\begin{align}
G^{\bold{k}}_{i\sigma,j\sigma'}(i\omega_{n}) = \int \frac{A^{\bold{k}}_{i\sigma,j\sigma'}(\omega)}{i\omega_{n} - \omega}d\omega . 
\end{align}
Due to the non-orthogonality of the GTO basis set, a basis transformation of the Green's functions to an orthonormal basis is necessary in order for $A^{\bold{k}}_{i\sigma,i\sigma}(\omega)$ to be normalized to one and strictly positive. 
In the present work, we choose symmetrized atomic orbitals (SAO)~\cite{Lowdin_sym_orth_1970} constructed from Gaussian Bloch orbitals. 
Expressing the Green's function in SAO, we continue all diagonal terms of the Green's function from the Matsubara frequency domain to the real frequency axis using Nevanlinna analytical continuation~\cite{Nevanlinna_Jiani_2021} to obtain the $\bold{k}$-resolved orbital-dependent spectral functions $A^{\bold{k}}_{i\sigma,i\sigma}(\omega)$, and the $\bold{k}$-resolved spectral functions $A^{\bold{k}}(\omega) = \sum_{i\sigma}A^{\bold{k}}_{i\sigma,i\sigma}(\omega)$. 
Note that for $\bold{k}$-points that are not included in the original $\bold{k}$-mesh, Wannier interpolation~\cite{Wannier_RMP_2021} is performed on self-energies in the GTO basis, and interpolated Green's functions are computed via the Dyson equation before continuation. 

\subsection{Complexity analysis\label{subsec:complexity}}
We now discuss the computational scaling and memory requirements of our sc$GW$ with density-decomposed interactions. 
All our calculations are performed in double precision. 
We define $N_{k}$ as the number of $\bold{k}$-points sampled in the Brillouin zone, $N_{orb}$ as the number of atomic orbitals in the unit cell, $N_{aux}$ as the number of auxiliary basis functions used to decompose two-electron integrals in the density-fitting procedure, and $N_{\tau}$ as the number of sparse sampling points on the imaginary-time axis. 
$N_{\tau}$ equals to the number of sampling points in the Matsubara domain ($N_{\omega}$). 

The calculation of the static Hartree-Fock part of the self-energy requires the evaluation of Eq.~\ref{Eq:HF_J} and Eq.~\ref{Eq:HF_K}. 
The evaluation of the Coulomb term scales as $\mathcal{O}(N_{k}N_{orb}^{2}N_{aux})$. 
The computational bottleneck at this step is the evaluation of the exchange potential $\bold{K}^{\bold{k}}$, which scales as $\mathcal{O}(N_{k}^{2}N_{orb}^{3}N_{aux})$.

The evaluation cost of the dynamical part of the self-energy, $(\tilde{\bold{\Sigma}}^{GW})^{\bold{k}}(\tau)$, is dominated by the complex dense matrix products in Eq.~\ref{Eq:tilde_P0_tau} and~\ref{Eq:tilde_sigma2} as well as the Dyson-like linear equation for $\tilde{\bold{P}}^{\bold{q}}(i\Omega_{n})$ in Eq.~\ref{Eq:tilde_P_omega}. 
The most costly steps in Eq.~\ref{Eq:tilde_P0_tau} and~\ref{Eq:tilde_sigma2} scale as $\mathcal{O}(N_{\tau}N_{k}^{2}N_{orb}^{2}N_{aux}^{2})$, while the linear equation can be solved in $\mathcal{O}(N_{\omega}N_{k}N_{aux}^{3})$. 

The evaluation of the linear system in Eq.~\ref{Eq:tilde_P_omega} has a low scaling with respect to $N_{k}$ and $N_{\omega}$. Inefficient sampling on the Matsubara axis, such as on a uniform Matsubara grid, will result in a large $N_{\omega}$ and a slow evaluation of Eq.~\ref{Eq:tilde_P_omega}. 
This bottleneck is avoided with sparse frequency sampling techniques~\cite{sparse_sampling_Jia_2020} (Sec.~\ref{subsec:IR}) for the bosonic functions $\tilde{\bold{P}}^{\bold{q}}_{0}(i\Omega_{n})$ and $\tilde{\bold{P}}^{\bold{q}}(i\Omega_{n})$. 
The resulting $N_{\omega}$ is reduced to 10 $\sim$ 200 sampling points, depending on the temperature and bandwidth of the system. 

Due to the lack of frequency dependence, the time to solution of the Hartree-Fock self-energies is typically $2\sim 3$ orders of magnitude smaller than the one of the dynamical self-energy part. 
The dynamical part is typically dominated by Eq.~\ref{Eq:tilde_P0_tau} and~\ref{Eq:tilde_sigma2}. 

The memory requirements are as follows. 
The largest objects in memory are the Green's function $G^{\bold{k}}_{i\sigma,j\sigma'}(\tau)$ and the self-energies $\Sigma^{\bold{k}}_{i\sigma,j\sigma'}(\tau)$, which scale as $\mathcal{O}(N_{\tau}N_{k}N_{orb}^{2})$. 
In our implementation, both objects are stored once per node in shared memory. 
The GDF integrals are precomputed and stored on disk. 
In the evaluation of the $GW$ self-energy, they are read in a batch of $\bold{k}$-points at a time, such that the memory requirement is $\mathcal{O}(N_{orb}^{2}N_{aux})$. 
The disk storage requirement scales as $\mathcal{O}(N_{k}^{2}N_{orb}^{2}N_{aux})$, and for large simulations, the storage needs may exceed 1 TB. 

\subsection{GPU acceleration\label{subsec:GPU_algorithm}}
\begin{algorithm}[H]\label{alg:GW_kernel}
\For{$\bold{q} \gets 1 \ \mathrm{to} \ N_{k}$}{
\For{$\bold{k} \gets 1 \ \mathrm{to} \ N_{k}$}{
Read GDF Coulomb tensors$(\bold{k},\bold{k+q})$\;
\For{$\tau \gets 1 \ \mathrm{to} \ N_{\tau}$}{
Eq.~\ref{Eq:tilde_P0_tau}$(\bold{q}, \bold{k}, \tau)$\;
}
}
Eq.~\ref{Eq:P0t_to_P0w}$(\bold{q})$\;
\For{$\Omega_{n}\gets 1 \ \mathrm{to} \ N_{\omega}$}{
Eq.~\ref{Eq:tilde_P_omega}$(\bold{q},\Omega_{n})$\;
}
Eq.~\ref{Eq:Piw_to_Pt}$(\bold{q})$\;
\For{$\bold{k} \gets 1 \ \mathrm{to} \ N_{k}$} {
Read GDF Coulomb tensors$(\bold{k},\bold{k-q})$\;
\For{$\tau \gets 1 \ \mathrm{to} \  N_{\tau}$}{
Eq.~\ref{Eq:tilde_sigma2}$(\bold{q},\bold{k},\tau)$\;
}
}
}
\caption{Pseudocode for the evaluation of the dynamical $GW$ self-energy $(\boldsymbol{\tilde{\Sigma}}^{GW})^{\bold{k}}(\tau)$.}
\end{algorithm}

The computational bottleneck of sc$GW$ equations is the evaluation of the dynamical $GW$ self-energy (the blue box in Fig.~\ref{Fig:scGW_workflow}). 
To facilitate its calculation, this part is accelerated using MPI and CUDA. 

Algorithm~\ref{alg:GW_kernel} shows a pseudocode for the evaluation of the dynamical $GW$ self-energy. 
A parenthesis indicates function arguments.  
A naive parallelization strategy would be a scheme that distributes the outermost loop over $\bold{q}$-points to different MPI processes and performs the calculations independently. 
However, such a scheme will only be able to reach peak performance for systems with extremely large sizes. 
To fully maximize the throughput of GPUs for arbitrary systems sizes, multiple layers of parallelization are needed. 

We first distribute small batches of $\bold{q}$-points to different MPI processes as the first layer of parallelization. Each process is assigned one GPU at which the modified bare polarization function $\boldsymbol{\tilde{P}}^{\bold{q}}_{0}$ (Eq.~\ref{Eq:tilde_P0_tau}), the modified dielectric function $\boldsymbol{\tilde{P}}^{\bold{q}}$ (Eq.~\ref{Eq:tilde_P_omega}), and the corresponding contributions to $(\boldsymbol{\tilde{\Sigma}}^{GW})^{\bold{k}}(\tau)$ (Eq.~\ref{Eq:tilde_sigma2}) for a given $\bold{q}$-batch are calculated. The size of a $\bold{q}$-batch will depend on the number of GPU cards available. 

Different $\bold{q}$-points in the same local $\bold{q}$-batch are processed serially. 
At this stage, the number of active MPI processes per node equals the number of available GPU per node. 
Within the two intermediate \texttt{for} loops over $\bold{k}$-points, multiple asynchronous streams are created over the $\bold{k}$-axis as a second layer of parallelization. 
This asynchronous stream handling allows overlaps between complex dense matrix multiplication (\texttt{ZGEMM}) with different $k$-indices for a given $\bold{q}$-point. The number of streams is determined automatically by the available GPU memory. 
Note that the parallelization at this layer may be inhibited by I/O operations, and the memory copying for the GDF Coulomb tensors $V^{\bold{k}\bold{k+q}}_{\ a \ b}(Q)$ between GPUs and CPUs. 
Asynchronous streams are used to hide this latency. 
When CPU memory is large enough to store V, reading the entire $V^{\bold{k}\bold{k+q}}_{\ a \ b}(Q)$ tensor at the beginning of the calculation is advantageous, such that the only overhead is the memory copy between CPUs and GPUs. 
Lastly, for loops over $N_{\tau}$ and $N_{\omega}$, we use batched versions of \texttt{ZGEMM}~\cite{MAGMA_batched_gemm} and a batched Cholesky linear system solver~\cite{MAGMA_batched_linear_solver} as a third layer of parallelization. The size of the $\tau$-batch is set as an external parameter to allow further fine-tuning. 

On top of the first layer of parallelization using MPI, the second and the third layer allow to fully utilize the computing resources on each GPU independent of system size. 
Once the computation of the local $GW$ self-energy is completed, data reductions for both $(\bold{\Sigma}^{GW}_{\infty})^{\bold{k}}$ and $(\tilde{\bold{\Sigma}}^{GW})^{\bold{k}}(\tau)$ are performed. 

\section{Results\label{sec:results}}
In this section, we list results from our sc$GW$ by analyzing sc$GW$ band gaps. In the present work, band gaps are defined as peak-to-peak distances in $\bold{k}$-resolved spectral functions, as defined in Sec.~\ref{subsec:spectral}. 
With Nevanlinna techniques~\cite{Nevanlinna_Jiani_2021}, such quantities can be evaluated accurately from imaginary time data. 
Except Sec.~\ref{subsec:results_thermodynamics}, the inverse temperature $\beta$ is always chosen to be 700 a.u. (corresponds to $\sim$ 450 K).  

\subsection{Validation of self-consistent solutions  \label{subsec:sc_validation}}
\begin{figure}[tbh!]
\includegraphics[width=0.45\textwidth]{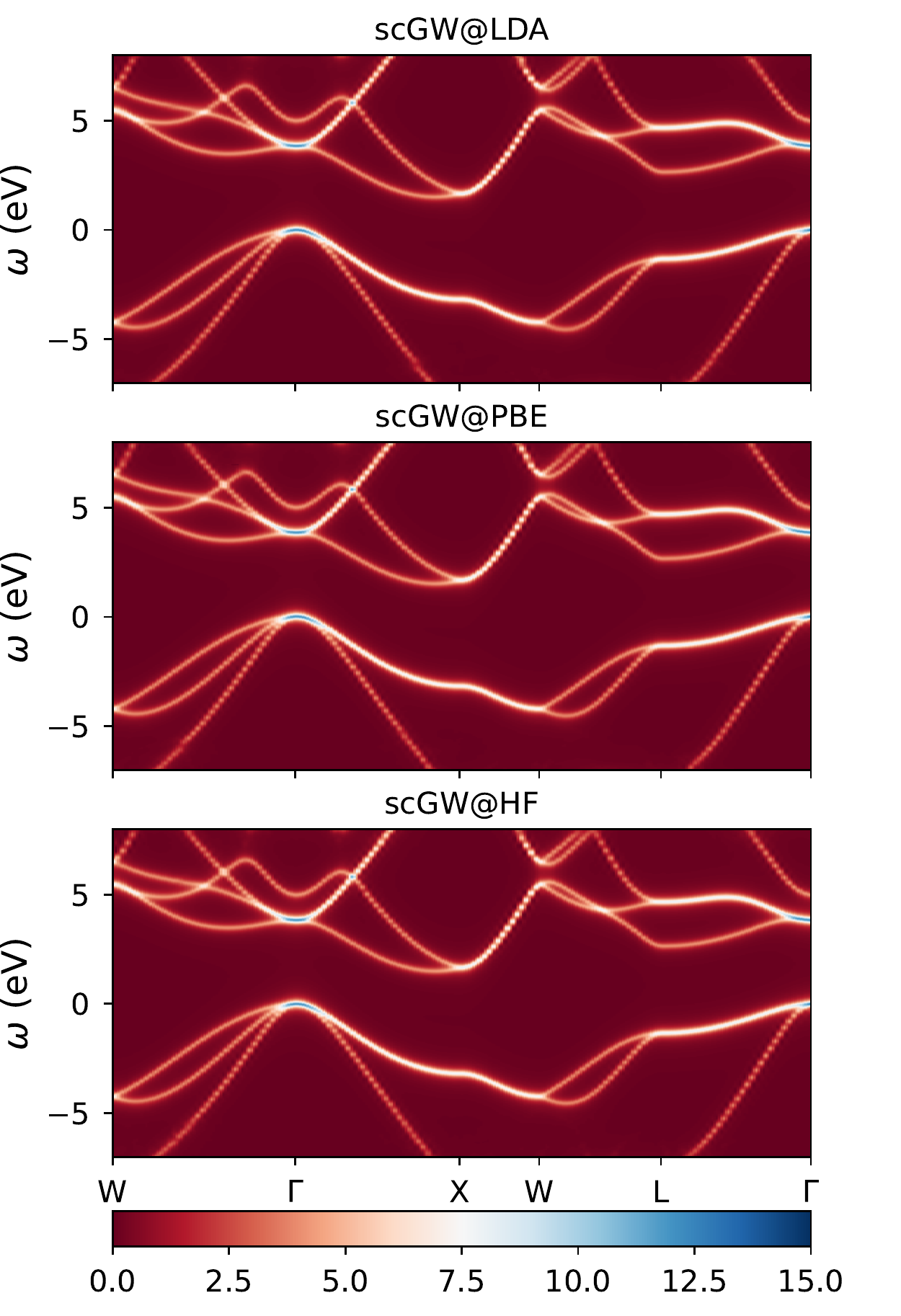}
\caption{sc$GW$ $\bold{k}$-resolved spectral functions of Si calculated from the LDA (top), PBE (middle), and HF (bottom) solutions.}\label{Fig:sc_validation}
\end{figure}

One of the main advantage of sc$GW$ is its independence of starting point on the final solution. 
Here, we verify this property by comparing the sc$GW$ results calculated from different starting solutions. 

Fig.~\ref{Fig:sc_validation} shows the sc$GW$ $\bold{k}$-resolved spectral functions of Si whose initial Green's functions are taken from LDA, PBE, and HF. 
Calculations are performed using a $6\times6\times6$ $\bold{k}$-mesh and the all-electron x2c-TZVPall basis set. 
The spectral functions are obtained via the prescription described in Sec.~\ref{subsec:spectral}. 
In spite of the different starting non-interacting Green's function, the sc$GW$ spectral functions converge to the same results consistently along the high-sysmmetry $\bold{k}$-path. 

\subsection{Thermodynamic consistency\label{subsec:results_thermodynamics}}
\begin{figure}[tb!]
\includegraphics[width=0.5\textwidth]{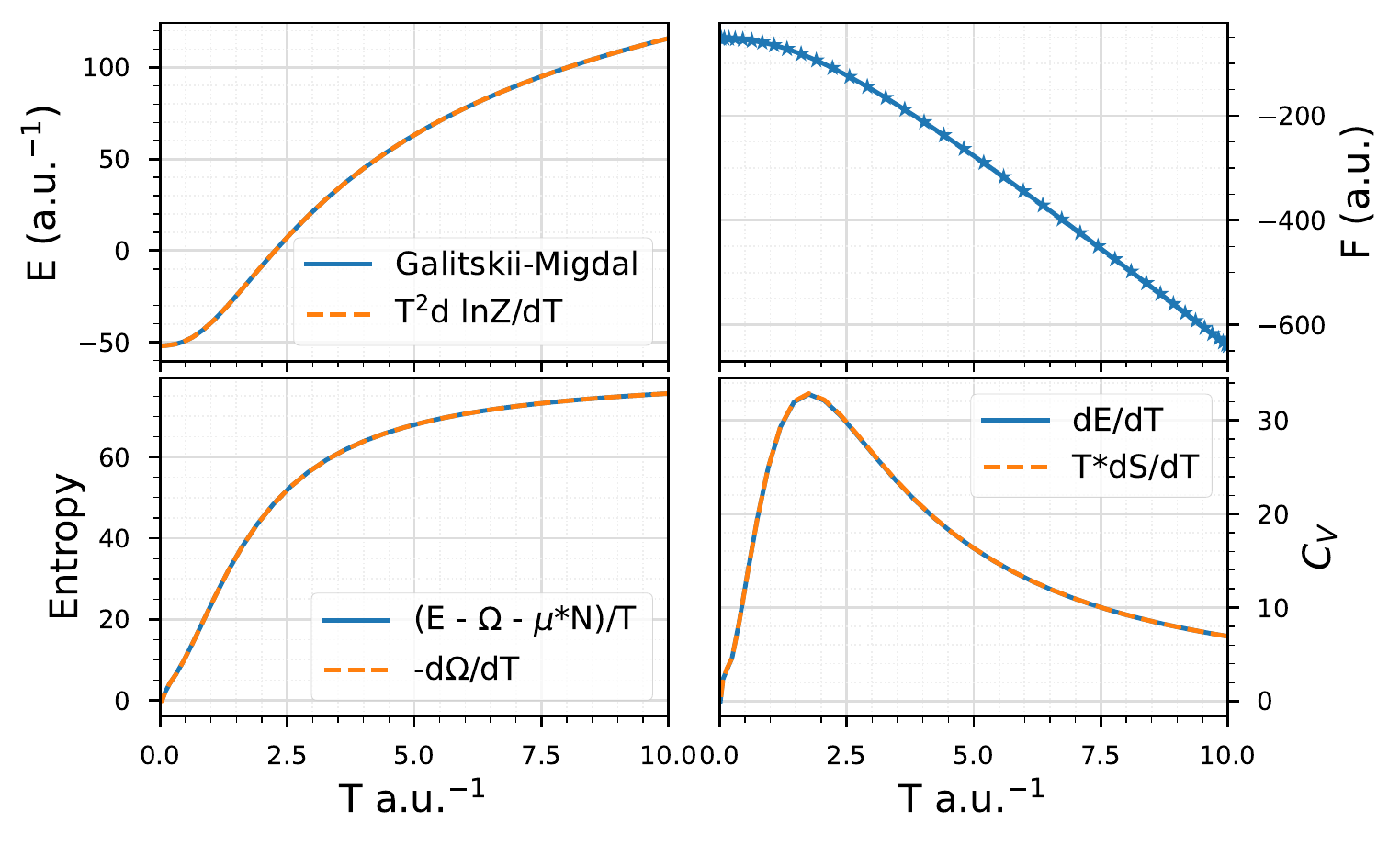}\\
\caption{Electronic thermodynamic quantities evaluated at fixed chemical potential $\mu=0$, including total energy ($E$), free energy ($F$), entropy ($S$), and specific heat ($C_{V}$), as functions of temperature for BN at $4\times4\times4$ $\bold{k}$-mesh.}\label{Fig:BN_thermodynamics}
\end{figure}

We continue our analysis of sc$GW$ by verifying its thermodynamic consistency.
Given self-consistent $GW$ solutions and the corresponding $\Phi$ functionals, we evaluate the electronic contribution to the thermodynamic quantities, including total energy ($E$), free energy ($F$), entropy ($S$), and specific heat ($C_{V}$), as functions of temperature. 
In Fig.~\ref{Fig:BN_thermodynamics}, we illustrate thermodynamic properties for BN calculated using a $4\times4\times4$ $\bold{k}$-mesh and the all-electron x2c-TZVPall basis set. 
Consistency between different ways of evaluating total energy, entropy, and specific heat verifies that sc$GW$ is thermodynamically consistent.

\subsection{Finite-size corrections\label{subsec:results_head_corr}} 
\begin{figure}[tbh!]
\includegraphics[width=0.45\textwidth]{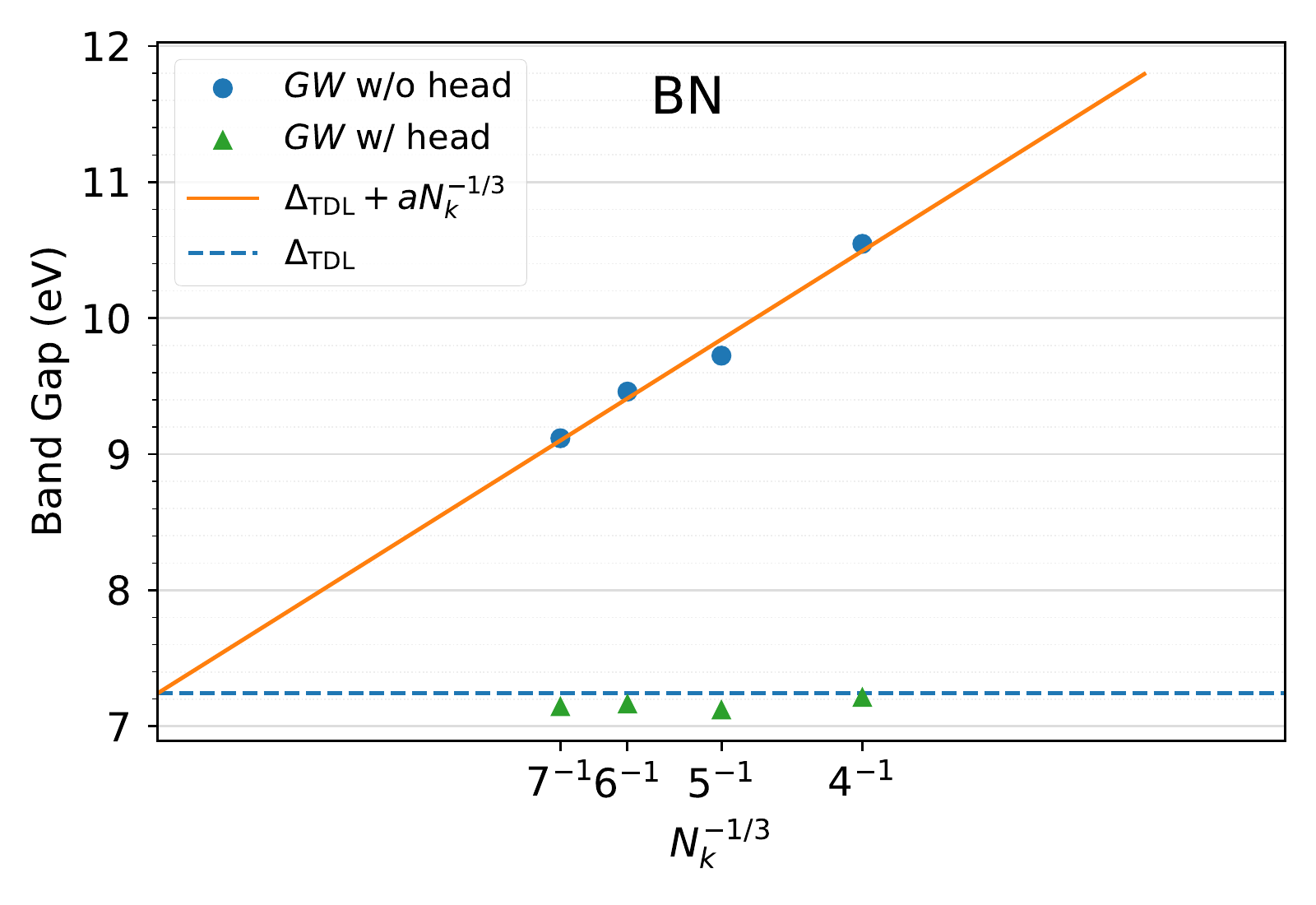}
\includegraphics[width=0.45\textwidth]{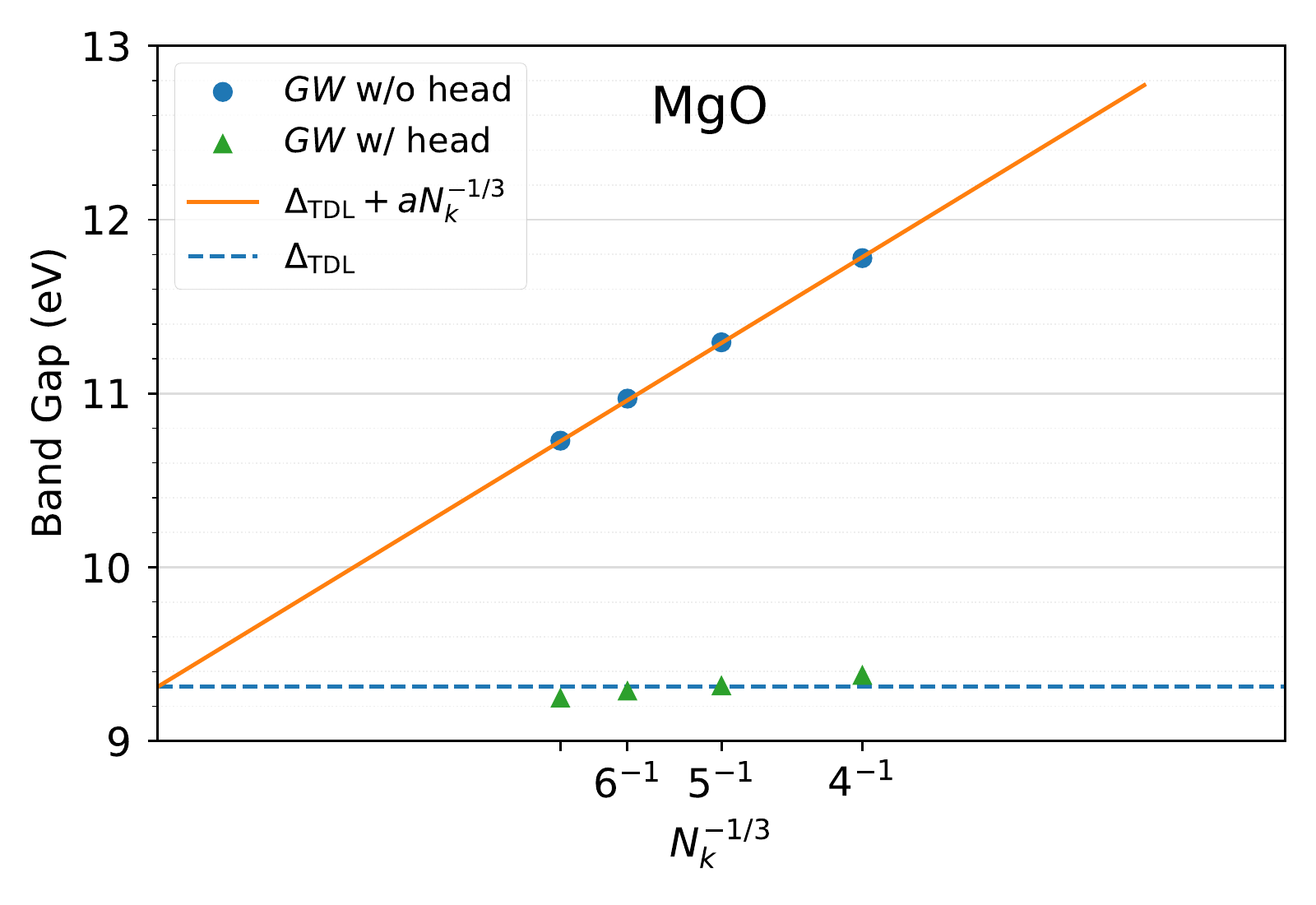}
\caption{sc$GW$ band gap as a function of $N_{k}^{-1/3}$ with and without the head corrections to the dynamical part of the $GW$ self-energy. A linear fit  is performed for the uncorrected band gaps (orange lines) to extrapolate the TDL values (blue dashed lines). }\label{Fig:TDL_conv}
\end{figure}

In this section, we analyze sc$GW$ by investigating the finite-size effects. 
By manually neglecting the singularity of the two-electron Coulomb integrals at $\bold{q} \rightarrow \bold{0}$ as discussed in Sec.~\ref{subsec:head_corr}, the integrable divergence is avoided. 
This will result in a slow convergence to the thermodynamic limit (TDL), scaling as $\mathcal{O}(N_{k}^{-1/3})$. 
We investigate the effect of applying the head corrections described in Sec.~\ref{subsec:head_corr} to the $GW$ screened exchange self-energy. 

Fig.~\ref{Fig:TDL_conv} shows the convergence of band gaps with and without the head corrections to the dynamical $GW$ self-energy $(\tilde{\bold{\Sigma}}^{GW})^{\bold{k}}(\tau)$ as a function of $N_{k}^{-1/3}$ for two systems, $\mathrm{BN}$ and $\mathrm{MgO}$. 
Note that the finite-size corrections to the HF exchange potential are always included. 

In the absence of the head corrections, sc$GW$ band gaps consistently exhibit a linear convergence with respect to $N_{k}^{-1/3}$. 
The band gap values are far from converged even with the largest $7\times7\times7$ $\bold{k}$-mesh. 
While the convergence is slow, band gaps in the thermodynamic limit can be extrapolated as demonstrated in Fig.~\ref{Fig:TDL_conv}. 
We perform the finite-size extrapolation by fitting the sc$GW$ band gap to $\Delta(N_{k}) = \Delta_{\mathrm{TDL}} + aN^{-1/3}_{k}$ for each system (orange lines) and extrapolate to the TDL value $\Delta_{\mathrm{TDL}}$ (blue dashed lines). 
The extrapolation yields the band gaps of $7.24$ eV for $\mathrm{BN}$ and $9.31$ eV for $\mathrm{MgO}$. 

When the head corrections are added to the dynamical sc$GW$ self-energy, a much faster convergence is observed consistently for all systems tested. 
A $4\times4\times4$ $\bold{k}$-mesh already results in band gap that is very close to the TDL value. 
The band gap values at the $7\times7\times7$ $\bold{k}$-mesh is 7.15 and 9.25 eV for BN and MgO which only differ with our extrapolated values by 0.1 eV. 
The same behavior is observed in all the test systems employed. 

Although an extrapolation to the TDL values for band gaps is possible, such a strategy may become impractical when quantities other than band gaps are of interest.
Different convergence patterns may be exhibited for these quantities. 

\subsection{Basis set convergence \label{subsec:basis_conv}}
\begin{table}[bth]
\begin{center}
\begin{tabular}{c|c|c|c}
\hline
\hline
Basis sets & x2c-SV(P)all & x2c-TZVPall  & x2c-QZVPall \\
\hline
Si & 1.87 & 1.54 & 1.55 \\
AlP & 2.97 & 2.90 & 2.96 \\
ZnO & 5.19 & 4.59 & 4.50 \\
ZnS & 4.87 & 4.58 & 4.46 \\ 
\hline
\hline
\end{tabular}
\caption{sc$GW$ band gaps (eV) of Si, AlP, ZnO, and ZnS calculated using different basis sets. A $5\times5\times5$ $\bold{k}$-mesh is used for Si and AlP, and a $4\times4\times4$ $\bold{k}$-mesh is used for ZnO and ZnS. In x2c-QZVPAll, the most diffuse $s$ and $p$ functions of Si, Al, and Zn are removed to avoid linear dependencies. 
\label{tab:basis_conv}}
\end{center}
\end{table}

In this section, we investigate the basis set convergence of sc$GW$ band gaps. 
Rather than employing non-relativistic calculations, we focus on the scalar relativistic case with spin-free X2C1e (sfX2C1e) Hamiltonian, since conventional non-relativistic GTO basis sets can become inadequate especially for the description of core electrons.  
We employ a family of all-electron basis sets optimized with the X2C Hamiltonian~\cite{X2CTZVP_Pollak_2017, X2CQZVP_Franzke}. 
The basis sets are systematically enlarged from a double-$\zeta$ (x2c-SV(P)all), to triple-$\zeta$ (x2c-TZVP), and finally to quadruple-$\zeta$ (x2c-QZVPall) basis set by adding additional high-lying atomic orbitals. 

Table~\ref{tab:basis_conv} shows the basis set convergence of sc$GW$ band gaps. 
The band gaps of Si converge very fast with the number of atomic basis functions. 
At x2c-TZVPall, the gaps are well converged, and only $\sim$ 0.01 eV difference is observed from x2c-TZVPall to x2c-QZVPall. 
The slightly slower convergence in AlP is likely due to the missing diffuse functions in x2c-QZVPall that are removed from Al to avoid linear dependencies. 
On the other hand, the convergence behavior in the presence of transition metal elements is the slowest as demonstrated by ZnO and ZnS. 
An additional band gap narrowing of about 0.1 eV is observed when going from x2c-TZVPall to x2c-QZVPall. 
Note that the slower convergence is well-known and is attributed to the $d$ orbitals of the transition metal elements. 
Similar behavior has also been observed in compounds with $4d$ transition elements, such as silver halides~\cite{GW_AgX_PRB_2018,AgX_LAPW_HLOs}. 

Overall, a large improvement is observed when going from a double-$\zeta$ basis to a triple-$\zeta$ basis. The quantitative differences between x2c-TZVPall and x2c-QZVPall are typically minor as long as no transition metal element is present. 
In the presence of transition metal elements, an additional band gap narrowing of $\sim 0.1$ eV is expected. 
Lastly, we have also investigated the basis convergence of the valence band maximum (VBM) and the conduction band minimum (CBM) in Appendix~\ref{appendix:basis_conv_for_bands}. 
We conclude that the convergence behavior observed in Table~\ref{tab:basis_conv} is not due to fortunate error cancellation between VBM and CBM.

\subsection{sc$GW$ band gaps\label{subsec:band_gap}}
\begin{figure}[tb!]
\begin{center}
\includegraphics[width=0.45\textwidth]{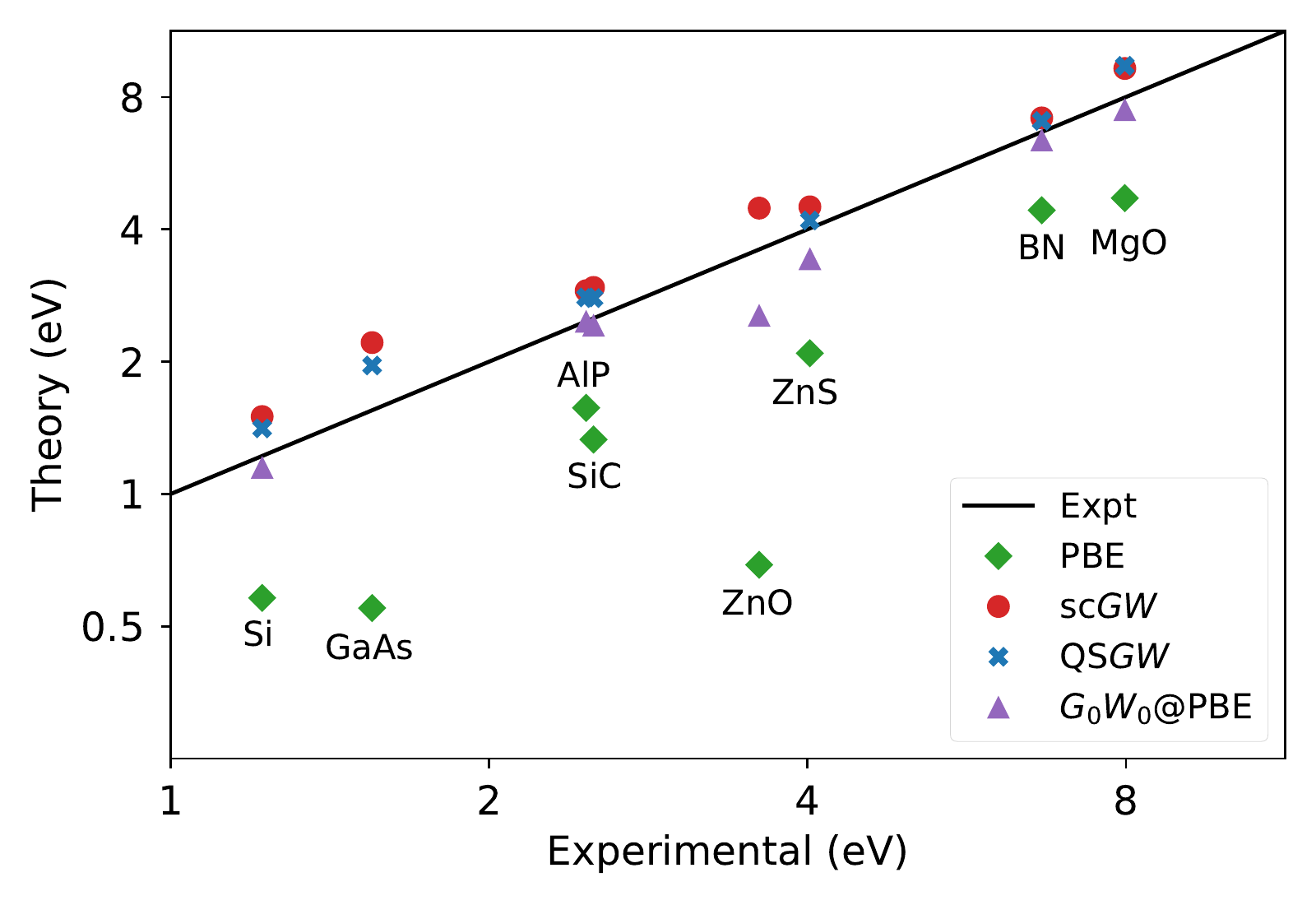}
\caption{Band gaps of selected semiconductors and insulators using all-electron sc$GW$ with sfX2C1e-Coulomb Hamiltonian in comparison with PBE functional and experimental data with corrections from ZPR as shown in Table~\ref{tab:all_electron_band_gaps}. The QS$GW$ band gaps taken from Ref.~\onlinecite{scGW_also_vertex_Andrey_2017} and the $G_{0}W_{0}$ (based on PBE functional) band gaps taken from Ref.~\onlinecite{scGW_VASP_2018} are also shown. }\label{Fig:band_gap_benchmark}
\end{center}
\end{figure}

\begin{table*}[bth]
\begin{center}
\begin{tabular}{c|c|ccc|cc|cc|cc}
\hline
\hline
System & PBE & & sc$GW$ & & QS$GW$ & & $G_{0}W_{0}$@PBE & & Expt & Expt+ZPR\\
& & This work & Ref.~\cite{scGW_also_vertex_Andrey_2017} & Ref.~\cite{scGW_VASP_2018} & Ref.~\cite{scGW_also_vertex_Andrey_2017} &  Ref.~\cite{scGW_VASP_2018} &Ref.~\cite{G0W0_Zhu2021} & Ref.~\cite{scGW_VASP_2018}\\ 
\hline
Si & 0.58 & 1.50 &1.55 & 2.18 &1.41 & 1.49 &1.08 & 1.15 &1.17~\cite{Kittel2004} & 1.22~\cite{ZPR_Si_SiC_2014}\\
SiC & 1.33 & 2.95 & 2.89 & 3.29 & 2.79 & 2.88 & 2.42 & 2.42 & 2.40~\cite{Yu2010} & 2.51~\cite{ZPR_Si_SiC_2014} \\
GaAs & 0.55 & 2.21 & 2.27 & & 1.96 &  & & &1.52~\cite{Kittel2004} & 1.55~\cite{ZPR_GaAs_2014} \\
AlP & 1.57 & 2.90 & 2.84 & 3.20 & 2.80 & 2.94 &  2.41& 2.47 & 2.45~\cite{Sze1981} & 2.47~\cite{ZPR_AlP_ZnO_ZnS_2005} \\
ZnO & 0.69 & 4.47 & & 4.92 & & 4.29 & 2.91 & 2.55 &3.44~\cite{Kittel2004} & 3.60~\cite{ZPR_AlP_ZnO_ZnS_2005} \\
ZnS & 2.09 & 4.50 & 4.28 & 4.68 & 4.19& 4.27 & 3.63& 3.43 &3.91~\cite{Kittel2004} & 4.02~\cite{ZPR_AlP_ZnO_ZnS_2005} \\ 
BN & 4.42 & 7.17 & 7.06 & 7.67 & 7.06 & 7.50 & 6.41 & 6.39 &6.40~\cite{madelung2004semiconductors} &6.66~\cite{ZPR_BN_MgO_2015} \\
MgO & 4.71 & 9.29 & 9.31 & 9.53 & 9.42 & 9.58 & 7.43 & 7.49 &7.83~\cite{MgO_Expt_1973} & 7.98~\cite{ZPR_BN_MgO_2015} \\
\hline
\hline
\end{tabular}
\caption{Band gaps (eV) of selected semiconductors and insulators calculated using all-electron sc$GW$ with sfX2C1e-Coulomb Hamiltonian in comparison with the experimental data~\cite{Kittel2004,Yu2010,Sze1981,madelung2004semiconductors,MgO_Expt_1973} with and without zero-point renormalization due to electron-phonon coupling~\cite{ZPR_Si_SiC_2014,ZPR_GaAs_2014,ZPR_AlP_ZnO_ZnS_2005,ZPR_BN_MgO_2015}. 
\label{tab:all_electron_band_gaps}}
\end{center}
\end{table*}

We now analyze our sc$GW$ by benchmarking the band gaps of systems for which experimental data exists. 
Table~\ref{tab:all_electron_band_gaps} shows the sc$GW$ band gaps of selected semiconductors and insulators calculated using sc$GW$ and DFT as well as theoretical~\cite{LQSGW_Kutepov2017,G0W0_Zhu2021} and experimental literature data~\cite{Kittel2004,Yu2010,Sze1981,madelung2004semiconductors,MgO_Expt_1973}. 
Zero-point renormalization (ZPR) due to electron-phonon coupling from existing calculations~\cite{ZPR_Si_SiC_2014,ZPR_GaAs_2014,ZPR_AlP_ZnO_ZnS_2005,ZPR_BN_MgO_2015} is taken into account with the raw experimental data to facilitate the comparison. 
For DFT calculations, the PBE density functional~\cite{PBE_Perdew_1996} is used. 
All calculations, including PBE, are based on all-electron sfX2C1e-Coulomb Hamiltonians with a $6\times6\times6$ $\bold{k}$-mesh. 
The sfX2C1e Hamiltonian incorporates the exact scalar relativistic effects at the one-electron level while neglecting the spin-orbit interactions and all the relativistic corrections to the electron-electron interactions. 
All our results, denoted as ``this work'', use all-electron triple-$\zeta$ bases optimized with respect to X2C Hamiltonians (x2c-TZVPall)~\cite{X2CTZVP_Pollak_2017}. 
For B and Mg atoms, the most diffuse $s$ and $p$ functions are removed to avoid linear dependencies. 

As shown in Table~\ref{tab:all_electron_band_gaps}, the PBE functional significantly underestimates the experimental band gaps. 
This trend has been observed in other works, see e.g.~\cite{scGW_VASP_2018,G0W0_Zhu2021}. 
The many-body treatment from sc$GW$ induces a gap widening.
The largest difference between PBE and sc$GW$ is observed for ZnO where the electron correlations from the transition metal $d$ orbitals are strong. 
While an overall good agreement  with experiment  is observed, especially when corrections from ZPR are taken into account, sc$GW$ systematically overestimates experimental band gaps as shown in Fig.~\ref{Fig:band_gap_benchmark}. 
Since there is no starting-point dependence in this self-consistent approximation, we argue that this overestimation is due to the absence of high-order diagrams, such as ``vertex corrections'' in the $GW$ self-energy and the polarizability that have been studied in the framework of bold diagrammatic expansions~\cite{GW_vertex_Gruneis_2014,scGW_w_vertex_Andrey_2016,scGW_also_vertex_Andrey_2017}. 
The effect of spin-orbit interaction is likely minor for the systems considered here. 
For instance, the strongest SOC effect is observed in GaAs, which exhibits a 0.1 eV band gap narrowing in Ref.~\onlinecite{QSGW_Chen2015} and in our in-house two-component sc$GW$ based on the the X2C1e-Coulomb Hamiltonian (not shown in the present work). 
Additional uncertainties, such as those caused by finite-size effects and basis set convergence are only expected to result in small quantitative differences, see Sec.~\ref{subsec:results_head_corr} and~\ref{subsec:basis_conv}. 

We compare our results to some of the $GW$ implementations~\cite{LQSGW_Kutepov2017,scGW_also_vertex_Andrey_2017,scGW_VASP_2018, G0W0_Zhu2021} available in the literature, including sc$GW$~\cite{scGW_also_vertex_Andrey_2017, scGW_VASP_2018}, quasiparticle $GW$ (QS$GW$)~\cite{LQSGW_Kutepov2017,scGW_VASP_2018}, and $G_{0}W_{0}$~\cite{scGW_VASP_2018, G0W0_Zhu2021} as shown in Table.~\ref{tab:all_electron_band_gaps}. 
Note that Ref.~\onlinecite{scGW_also_vertex_Andrey_2017} is chosen since, to the best of our knowledge, it is the only fully self-consistent finite-temperature $GW$ capable of calculating realistic solids in the LAPW basis.
The basis set therefore marks the only major difference between the methodology of Ref.~\onlinecite{scGW_also_vertex_Andrey_2017} and this work. 
The VASP implementation~\cite{scGW_VASP_2018} is chosen because different variants of $GW$ are reported in this work employing a projector augmented wave (PAW) basis. 
Calculations in Refs.~\onlinecite{LQSGW_Kutepov2017,scGW_also_vertex_Andrey_2017,scGW_VASP_2018, G0W0_Zhu2021} are performed using a $6\times6\times6$ $\bold{k}$-mesh (with the exception of $8\times8\times8$ $\bold{k}$-mesh for Si in Ref.~\onlinecite{scGW_VASP_2018}). 

In general, good agreement is reached between our data and the sc$GW$ data from Ref.~\onlinecite{scGW_also_vertex_Andrey_2017}. 
This is somewhat expected, since both the implementations are based on finite-temperature Green's function methods and executed on the imaginary axes exclusively. 
Both use no analytical continuation during the self-consistency loop, treat all electrons explicitly without the use of pseudopotentials, and define band gaps as the peak-to-peak distance of the spectral function. 
We attribute the main difference to Ref.~\cite{scGW_also_vertex_Andrey_2017} to the difference between the LAPW and GTO basis sets. 
Remaining small differences may therefore be attributed to  finite-size effects, which include the treatment of the integral divergence, the basis set error, and  uncertainty in the analytical continuation procedure. 
For example, the larger deviations observed for ZnO and ZnS are consistent with the slower basis convergence for Zn atom as discussed in Sec.~\ref{subsec:basis_conv}. 

The comparison to VASP~\cite{scGW_VASP_2018} is somewhat surprising. 
Overall, we found that the sc$GW$ band gaps from VASP are generally larger than ours and those in Ref.~\onlinecite{scGW_also_vertex_Andrey_2017}. 
Even for a simple system such as silicon, a 0.65 eV larger band gap is observed in Ref.~\onlinecite{scGW_VASP_2018}. 
Several aspects may be responsible for these differences. 
Numerically, VASP uses a PAW basis, which implies different basis set convergence, and a different treatment of relativistic effects. 
In addition, the band gaps in VASP are defined as the quasiparticle gaps (evaluated in the last step of the algorithm). 
Given a self-energy calculated from sc$GW$ without a quasiparticle approximation, the band gaps are determined by solving a quasiparticle equation in the HF canonical-orbital basis in the post-processing step. 
In that case, all off-diagonal self-energies in the HF canonical-orbital basis are neglected, potentially resulting in overestimation of the gaps similar to the one observed in Ref.~\onlinecite{Caratheodory_Jiani_2021}. 
Another difference comes from the fact that the sc$GW$ in VASP is the zero-temperature version unlike ours and the one in Ref.~\onlinecite{scGW_also_vertex_Andrey_2017}. 
However, this difference should be negligible since the thermal excitations at the temperature ($\beta = 700$ a.u.) used here are not expected to affect the resulting spectral functions for systems studied in the present work. 

As for the two selected $G_{0}W_{0}$ results~\cite{scGW_VASP_2018,G0W0_Zhu2021}, a band-gap narrowing is observed in comparison to our sc$GW$ results. 
Note that larger deviations between Refs.~\onlinecite{G0W0_Zhu2021,scGW_VASP_2018} appear in ZnO and ZnS. 
As Ref.~\onlinecite{G0W0_Zhu2021} suggests, this is possibly due to different treatments for the core electrons. 
In calculations from Ref.~\onlinecite{G0W0_Zhu2021}, using all-electron GTO basis, the relativistic effects are completely ignored. 
Therefore, an additional band narrowing due to the scalar relativistic effect is expected. 
Note that, in spite of the numerical similarities due to the use of GTO basis sets, the core electrons are accurately treated with the scalar relativistic effects in the sfX2C1e Hamiltonian that is employed in our implementation. 

Overall, as expected, sc$GW$ results in larger band gaps when compared to $G_{0}W_{0}$. 
In $G_{0}W_{0}$, bandgaps for semiconductors in the absence of transition metal elements are expected to be slightly underestimated, due to an error cancellation between the lack of self-consistency and the vertex corrections~\cite{GW_vertex_Gruneis_2014,scGW_also_vertex_Andrey_2017}. 
Such an error cancellation is missing both in sc$GW$ and QS$GW$, and results in systematic overestimation of the band gap. 
With our data, we cannot confirm the observation of Ref.~\onlinecite{scGW_VASP_2018} that the overestimation is generally larger in sc$GW$ than QS$GW$. 
Lastly, the good agreement between our results and those in Ref.~\onlinecite{scGW_also_vertex_Andrey_2017} suggests that consistent and reproducible sc$GW$ results independent of the basis set employed are now possible. 

\subsection{GPU performance \label{subsec:GPU_profile}}
\begin{figure*}[tb!]
\includegraphics[width=0.325\textwidth]{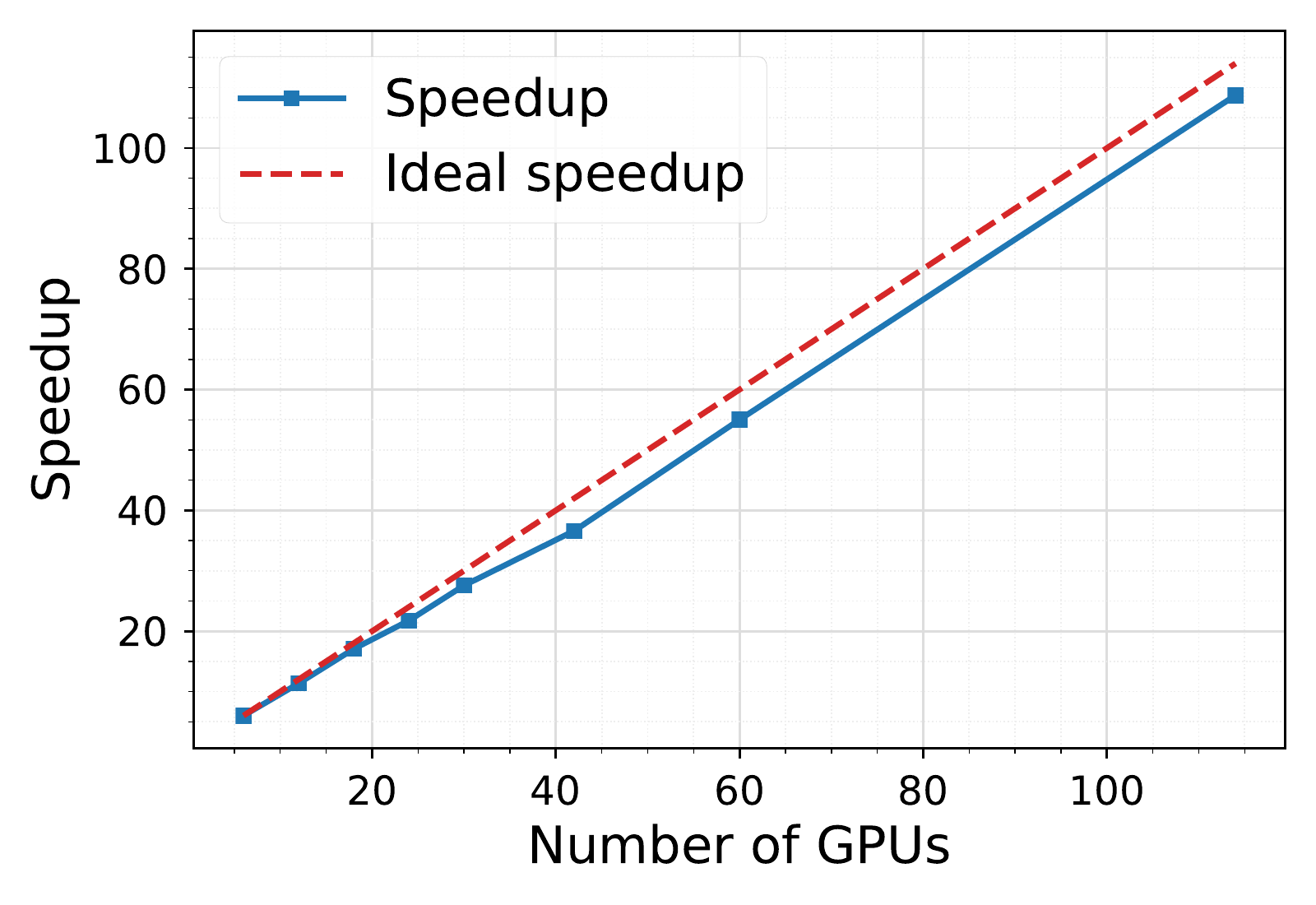}
\includegraphics[width=0.325\textwidth]{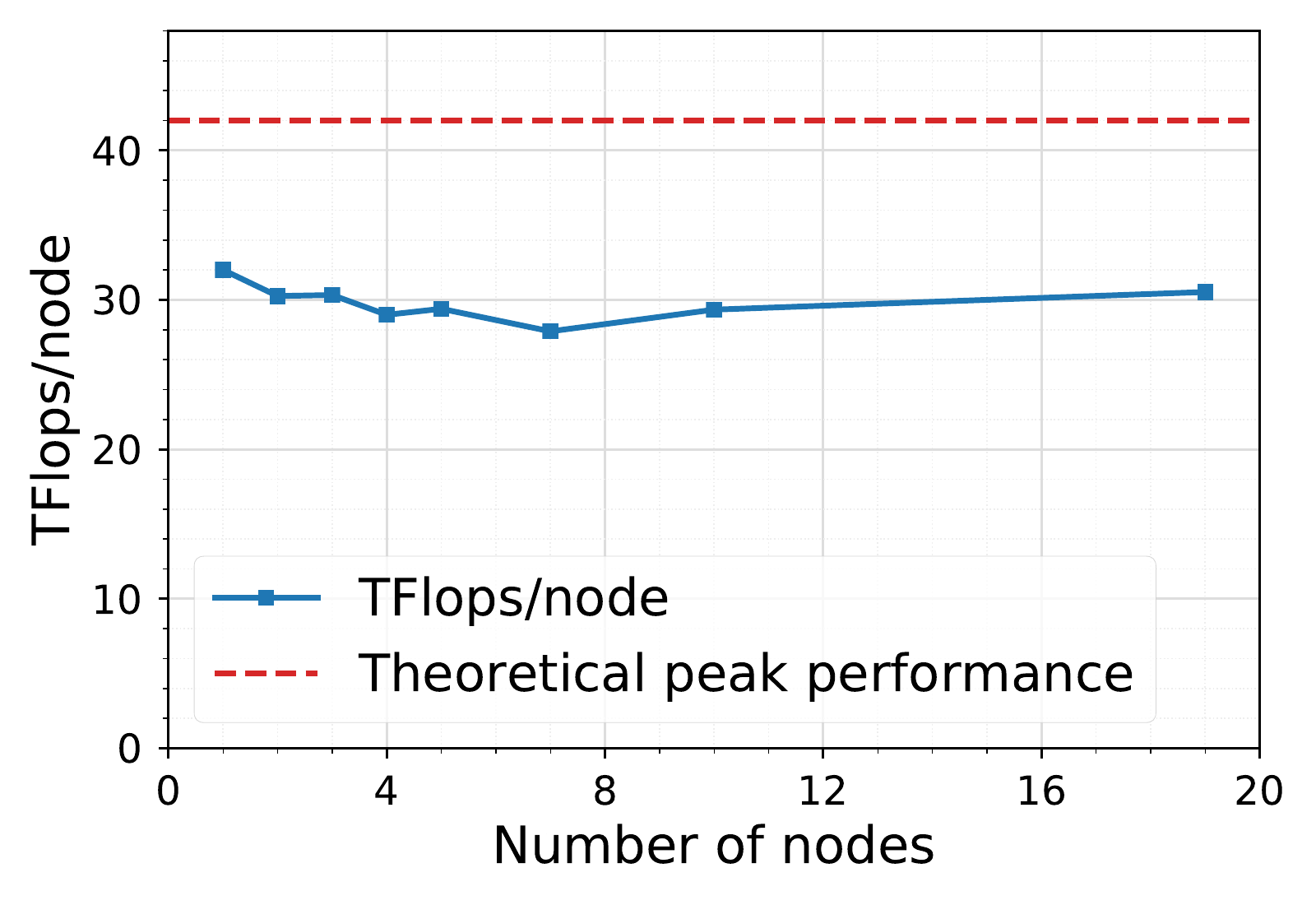}
\includegraphics[width=0.325\textwidth]{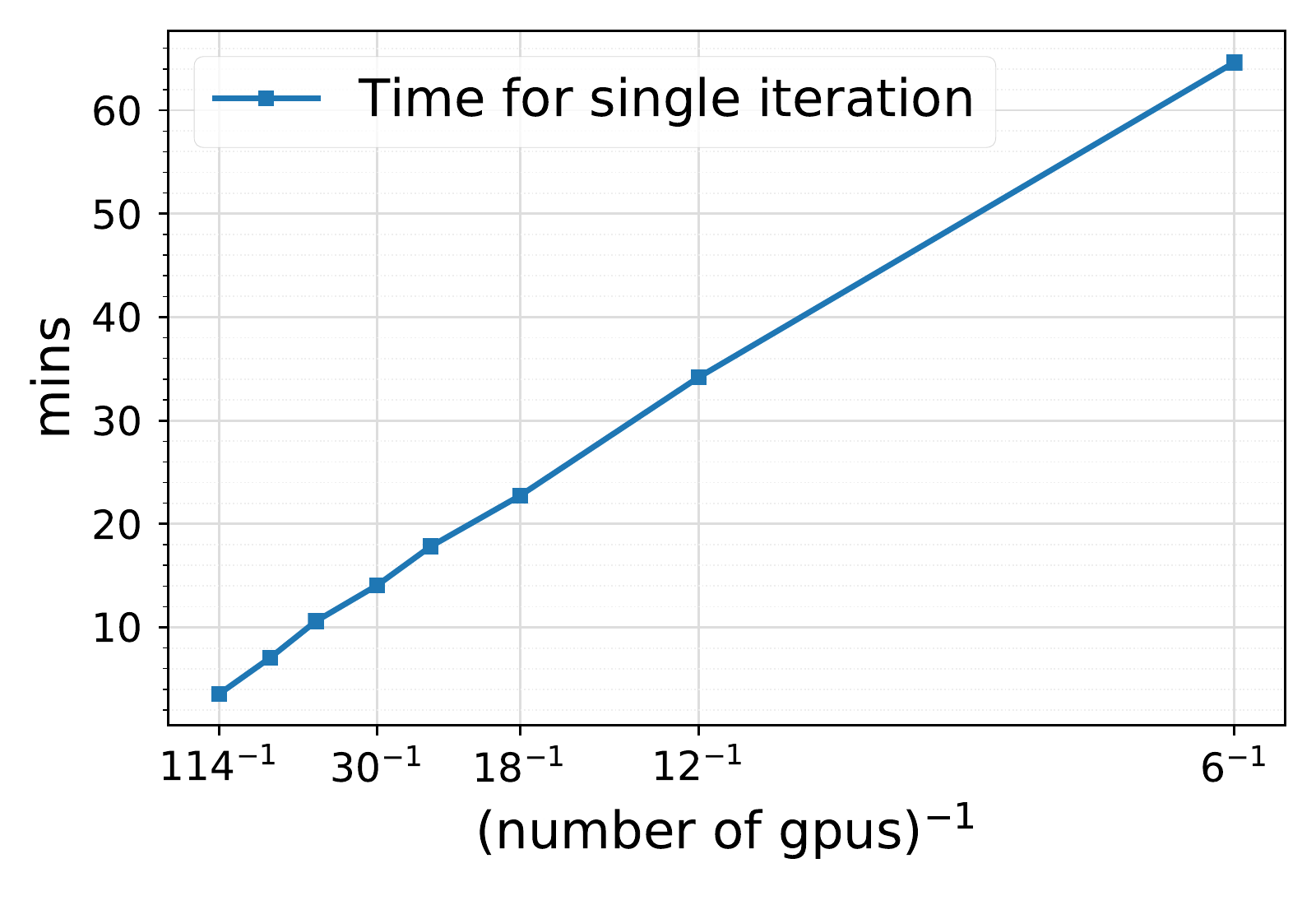}
\caption{Profiles of our GPU kernels for the evaluation of $(\tilde{\bold{\Sigma}}^{GW})^{\bold{k}}(\tau)$.}\label{Fig:GPU_profiles}
\end{figure*}

In this section, we demonstrate the performance of our implementation of sc$GW$. We will focus in particular on GPU kernels for the evaluation of the dynamical part of $GW$ self-energy $(\tilde{\Sigma}^{GW})^{\bold{k}}_{ij}(\tau)$ which is the computational bottleneck in our implementation. 

Fig.~\ref{Fig:GPU_profiles} shows a profiling result of our GPU kernels on \texttt{summit} at the Oak Ridge Leadership Computing Facility. 
Each node consists of six Nvidia Volta V100s GPU cards. 
Antiferromagnetic (AFM) MnO is chosen as the test system with a $6\times 6\times 6$ $\bold{k}$-mesh centered at the $\Gamma$-point, using the \emph{gth-dzvp-molopt-sr} basis~\cite{GTH_basis_2007} and the \emph{gth-pbe} pseudopotential~\cite{GTH_pseudo_1996}. 
Considering only inversion symmetry, the number of effective $\bold{q}$-points is 112. 
As shown in the first panel, the implementation exhibits almost ideal speedup consistently from a single GPU to 112 GPUs. 
The middle panel shows that around 70\% of the theoretical peak performance is achieved consistently on up to 20 nodes.  
The third panel shows the total time of evaluating $(\tilde{\bold{\Sigma}}^{GW})^{\bold{k}}(\tau)$ per iteration. 
This value includes communication and I/O overhead. 
When the first layer of MPI parallelization over the $\bold{q}$-axis is fully exploited, one iteration of the $GW$ self-energy evaluation takes $\sim3$ minutes for this particular example. 

\section{Conclusion \label{sec:conclusion}}
In this paper, we present implementation details and results for a fully self-consistent finite-temperature $GW$ method in Gaussian Bloch orbitals for solids. 
The method employs finite-temperature Green's function on the imaginary axis. 
The full self-consistency between Green's functions and self-energies guarantees that results are conserving and thermodynamically consistent. 
We do not employ the quasiparticle approximation, and all matrix elements of the $GW$ self-energy at all Matsubara frequencies are evaluated explicitly. 
Instead of calculating a quasiparticle gap, single-particle excitation information is obtained directly from a spectral function, calculated using Green's function analytically continued from the  imaginary to real frequency axis. 

The finite-temperature self-consistent $GW$ is computationally feasible due to various numerical developments employed in the present work. 
In particular, Gaussian density fitting reduces the complexity of the sc$GW$ algorithm to $\mathcal{O}(N_{\tau}N_{k}^{2}N_{orb}^{2}N_{aux}^{2})$. 
A compact representation of dynamical quantities using sparse sampling on imaginary axis with IR basis greatly reduces the memory requirement for dynamical quantities. 
More importantly, computational overheads of the Dyson-like equation for the bosonic function (the renormalized auxiliary function $\tilde{\bold{P}}^{\bold{q}}(i\Omega_{n})$ in our case), as well as frequency integration along Matsubara axis, and Fourier transformation between two imaginary axes are negligible, compared to the evaluation of self-energy and polarization function (or the non-interacting auxiliary function $\tilde{\bold{P}}^{\bold{q}}_{0}$ in our case). 

Moreover, we explore additional acceleration of the sc$GW$ algorithm by migrating computationally intensive parts to a hybrid CPU/GPU platform. 
We demonstrate that this implementation scales to hundreds of GPUs, with good scalability on large  systems. 

Lastly, the Nevanlinna analytical continuation as a post processing step makes the execution of $GW$ exclusively on imaginary axis possible,  by providing access to causal high-quality real-frequency data.
We note that we did not yet explore additional optimizations, such as those based on the locality of the self-energy and basis functions~\cite{cubicGW_spatialFFT_VASP_Kaltak2014,LowScaling_G0W0_CP2K_2018,linear_scGW_Kutepove_2021,G0W0_truncated_C_GDF_CP2K_2021}, and optimum basis sets for Gaussian Bloch orbitals~\cite{Yanbing_basis_opt,GTH_basis_Ye_2021}
as well as auxiliary bases in periodic systems. 
These are interesting options for future development that will facilitate large-unit-cell calculations for sc$GW$. 

In our analysis of sc$GW$, we demonstrate its thermodynamic consistency in practice, investigate the finite size effects with and without the head correction to the dynamical $GW$ self-energy, and investigate the basis convergence of sc$GW$ in Gaussian Bloch orbitals. 
Our benchmark employing band gaps of selected semiconductors and insulators shows consistent results when compared to the finite-temperature sc$GW$ implementation reported in Ref.~\onlinecite{scGW_also_vertex_Andrey_2017}, where the numerical setup is substantially different. 
This agreement shows that sc$GW$  it is now routinely possible to reach high quality results that are converged with respect to the basis set and finite size effects.

Without a quasiparticle approximation and with inclusion of the full  self-consistency, our work provides a direct assessment of the fully self-consistent $GW$ method when applied to realistic materials. 
Deviations from experimental data can be attributed to higher order self-energy diagrams. These diagrams can be added either by employing vertex corrections~\cite{GW_vertex_Gruneis_2014,QSGW_w_vertex_Kresse_2007,scGW_also_vertex_Andrey_2017} to the $GW$ self-energy diagrams or by using embedding methods~\cite{DMFT_RMP_1996,electronic_strcuture_w_DMFT_RMP_2006,Dominika_SEET_2017,Alexie_SEET_PRB2015,SEET_NiO_MnO_2020,SEET_perovskites_Yeh_2021,SEET_GFCCSD_Yeh_2021,Nilsson17,GW_EDMFT_PRM_Philipp20,Boehnke16} on top of sc$GW$. 

\section*{Acknowledgements}
This research used resources of the Oak Ridge Leadership Computing Facility at the Oak Ridge National Laboratory, which is supported by the Office of Science of the U.S. Department of Energy under Contract No. DE-AC05-00OR22725.
D.Z. and Ch.-N. Y. were supported by a grant from the Department of Energy under Award Number DE‐SC0022198.
This material is based upon work supported by the U.S. Department of Energy, Office of Science, Office of Advanced Scientific Computing Research and Office of Basic Energy Sciences, Scientific Discovery through Advanced Computing (SciDAC) program under Award Number DE‐SC0022198.
E.G and S.I. were supported by the Simons Foundation via the Simons Collaboration on the Many-Electron problem.

\bibliographystyle{apsrev4-2}
\bibliography{scGW_refs}

\appendix
\section{Integrable divergence treatment \label{appendix:div_treatment}}
In this section, we follow the procedure described in Ref.~\onlinecite{ERI_correction_2009} and derive the finite-size corrections for both the HF exchange potential (Eq.~\ref{Eq:K_corr}) and the dynamical $GW$ self-energy (Eq.~\ref{Eq:GW_head_corr}) as shown in Sec.~\ref{subsec:head_corr}. 

Considering a general numerical problem that involves an integral over the first Brillouin zone whose integrand contains a smooth function $A$ and the bare Coulomb kernel expressed in the plane-wave basis ($\bold{G}$), 
\begin{align}
X = \frac{-1}{(2\pi)^{3}}\int_{\mathrm{BZ}} d\bold{q}\sum_{\bold{G}}\frac{4\pi}{|\bold{q+G}|^{2}}A(\bold{q},\bold{G}).
\label{Eq:problem_X}
\end{align}
Analytically, the integral is integrable although the integrand diverges as $1/\bold{q}^{2}$ at $\bold{G=0}$ when $\bold{q}\rightarrow\bold{0}$. 
However, this singularity forbids a direct numerical evaluation using discretized $\bold{q}$-mesh. 
The numerical evaluation of Eq.~\ref{Eq:problem_X} directly resembles the evaluations of the HF exchange potential (Eq.~\ref{Eq:HF_K}) and the dynamical $GW$ self-energy (Eq.~\ref{Eq:tilde_sigma}). 
We will show how to solve Eq.~\ref{Eq:problem_X} numerically and apply the same strategy to Eq.~\ref{Eq:HF_K} and~\ref{Eq:tilde_sigma}. 

We subtract and add the integrand of Eq.~\ref{Eq:problem_X} by an auxiliary function $F(\bold{q},\bold{G})$ that exhibits the same divergence $\sim 1/\bold{q}^{2}$ as $\bold{q}\rightarrow\bold{0}$ at $\bold{G=0}$, i.e. 
\begin{align}
X = \frac{-1}{2\pi^{2}}&\int_{\mathrm{BZ}} d\bold{q}\sum_{\bold{G}}\bigg\{\frac{1}{|\bold{q+G}|^{2}}A(\bold{q},\bold{G}) - F(\bold{q},\bold{G})A(\bold{0},\bold{0})\bigg\} \nonumber \\
&+ \frac{-1}{2\pi^{2}}\int_{\mathrm{BZ}} d\bold{q}\sum_{\bold{G}}F(\bold{q},\bold{G})A(\bold{0},\bold{0}).  
\label{Eq:problem_X2}
\end{align}
In the long-wavelength limit ($\bold{q}\rightarrow\bold{0}$), the singularity of bare Coulomb kernel  is cancelled by the one of the auxiliary function. 
The resulting smooth integrand in the curly brackets can therefore be evaluated accurately by a summation over  a finite number of $\bold{q}$-points. 
On the other hand, the singularity has been transferred to the auxiliary function $F(\bold{q},\bold{G})$ in the last term in Eq.~\ref{Eq:problem_X2} which can be evaluated analytically. 

Approximating the integral by a discrete summation $\frac{1}{(2\pi)^{3}}\int_{\mathrm{BZ}} d\bold{q}\rightarrow \frac{1}{\Omega N_{k}}\sum_{\bold{q}}$ and rearranging different terms, Eq.~\ref{Eq:problem_X2} can be expressed as 
\begin{align}
X \approx -\sum_{\bold{q}}\sum_{\bold{G}}\Phi(\bold{q},\bold{G})A(\bold{q},\bold{G})
\label{Eq:problem_X3}
\end{align} 
where 
\begin{align}
\Phi(\bold{q},\bold{G}) = 
\begin{cases}
\chi \ \mathrm{for}\  \bold{q=G=0},\\ 
\frac{1}{N_{k}\Omega}\frac{4\pi}{|\bold{q+G}|^{2}}\ \mathrm{otherwise},
\end{cases}
\end{align}
with 
\begin{subequations}
\begin{align}
\chi &= \frac{1}{2\pi^{2}}\int_{\mathrm{BZ}} d\bold{q}\sum_{\bold{G}}F(\bold{q},\bold{G}) - \frac{4\pi}{\Omega N_{k}}\sum_{\bold{q}}\sum^{_{'}}_{\bold{G}}F(\bold{q},\bold{G})\\
&=\frac{1}{2\pi^{2}}\int d\bold{Q} F(\bold{Q}) - \frac{4\pi}{\Omega N_{k}}\sum_{\bold{Q}\neq\bold{0}}F(\bold{Q}). 
\end{align}
\end{subequations}
The summation with the prime symbol implies $\bold{G=0}$ is not included when $\bold{q=0}$. In the second line, we define $\bold{Q = q + G}$. 
The singularity at $\bold{G=q=0}$ is included in $\chi$ which is properly treated through analytical integration. 
The choice of $F(\bold{Q})$ will affect the smoothness of the integrand in the parentheses in Eq.~\ref{Eq:problem_X2}, and therefore affect the convergence with respect to the number of number of $\bold{q}$-points to approximate the integral as shown in Eq.~\ref{Eq:problem_X3}. 
In the present work, the auxiliary function proposed in Ref.~\onlinecite{ERI_correction_2009} is adopted which makes $\chi$ the supercell Madelung constant. 

Both the HF exchange potential and the dynamical $GW$ self-energy can be written in a similar format as in Eq.~\ref{Eq:problem_X}. 
The HF exchange potential in the plane-wave basis reads, 
\begin{align}
K^{\bold{k}}_{i\sigma,j\sigma'} = \frac{-1}{(2\pi)^{3}}\int_{\mathrm{BZ}}&d\bold{q}\sum_{\bold{G}}\sum_{ab}\frac{4\pi}{|\bold{q+G}|^{2}}\nonumber\\
&\times\rho^{\bold{k-q}\bold{k}*}_{\ \ a \ \ i}(\bold{G})\gamma^{\bold{k-q}}_{a\sigma,b\sigma'}\rho^{\bold{k-q}\bold{k}}_{\ \ b \ \ j}(\bold{G})
\end{align}
with
\begin{align}
A(\bold{q},\bold{G}; \bold{k}&,i\sigma, j\sigma'
) \nonumber\\
&= \sum_{ab}\rho^{\bold{k-q}\bold{k}*}_{\ \ a \ \ i}(\bold{G})\gamma^{\bold{k-q}}_{a\sigma,b\sigma'}\rho^{\bold{k-q}\bold{k}}_{\ \ b \ \ j}(\bold{G}). 
\end{align}
The corresponding finite-size correction reads
\begin{align}
(\Delta^{\mathrm{HF}})^{\bold{k}}_{i\sigma,j\sigma'} = -\chi \sum_{ab}S^{\bold{k}}_{ia}\gamma^{\bold{k}}_{a\sigma,b\sigma'}S^{\bold{k}}_{bj}. 
\end{align}
Note that 
\begin{align}
\rho^{\bold{k-q}\bold{k}}_{\ \ i\  \ j}(\bold{G})\Big|_{\bold{q=G=0}} &= \int_{\Omega}d\bold{r} \rho^{\bold{k-q}\bold{k}}_{\ \ i \ \ j}(\bold{r}) e^{-i(\bold{q}+\bold{G})}\Big|_{\bold{q=G=0}} \nonumber\\
&= \int_{\Omega} d\bold{r}\rho^{\bold{k}\bold{k}}_{i j} = S^{\bold{k}}_{ij}.
\end{align}
Similarly, we express the dynamical $GW$ self-energy in the plane-wave basis,
\begin{align}
(&\tilde{\Sigma}^{GW})^{\bold{k}}_{i\sigma,j\sigma'}(\tau)  = \frac{-1}{(2\pi)^{3}}\int_{\mathrm{BZ}}d\bold{q}\sum_{\bold{G}\bold{G}'}\sum_{ab} G^{\bold{k-q}}_{a\sigma,b\sigma'}(\tau) \\ 
&\rho^{\bold{k-q}\bold{k}*}_{\ \ a \ \ i}(\bold{G})\frac{\sqrt{4\pi}}{|\bold{q}+\bold{G}|}(\epsilon^{\bold{q},-1}_{\bold{G}\bold{G}'}(\tau) - \delta_{\bold{G}\bold{G}'})\frac{\sqrt{4\pi}}{|\bold{q}+\bold{G}'|}\rho^{\bold{k-q}\bold{k}}_{\ \ b \ \ j}(\bold{G}'). \nonumber
\end{align}
Since we are only interested in the correction to the head of $(\tilde{\Sigma}^{GW})^{\bold{k}}$ which corresponds to $\bold{G}=\bold{G}'=\bold{0}$, we consider only the diagonal terms in the plane-wave basis, i.e. $\bold{G}=\bold{G}'$, 
\begin{align}
(&\tilde{\Sigma}^{GW}_{\mathrm{diag}})^{\bold{k}}_{i\sigma,j\sigma'}(\tau)  = \frac{-1}{(2\pi)^{3}}\int_{\mathrm{BZ}}d\bold{q}\sum_{\bold{G}}\sum_{ab} \frac{4\pi}{|\bold{q}+\bold{G}|^{2}} \\ 
&(\epsilon^{\bold{q},-1}_{\bold{G}\bold{G}}(\tau) - 1)\rho^{\bold{k-q}\bold{k}*}_{\ \ a \ \ i}(\bold{G})G^{\bold{k-q}}_{a\sigma,b\sigma'}(\tau)\rho^{\bold{k-q}\bold{k}}_{\ \ b \ \ j}(\bold{G}), \nonumber
\end{align}
with 
\begin{align}
A(\bold{q},\bold{G}; \bold{k},i\sigma, j\sigma'&, \tau) = \sum_{ab}(\epsilon^{\bold{q},-1}_{\bold{G}\bold{G}}(\tau) - 1)\nonumber \\
&\rho^{\bold{k-q}\bold{k}*}_{\ \ a \ \ i}(\bold{G})G^{\bold{k-q}}_{a\sigma,b\sigma'}(\tau)\rho^{\bold{k-q}\bold{k}}_{\ \ b \ \ j}(\bold{G}).
\end{align}
The head correction then reads
\begin{align}
(\Delta^{GW})^{\bold{k}}_{i\sigma,j\sigma'}(\tau) = -\chi (\epsilon^{\bold{0},-1}_{\bold{0}\bold{0}}(\tau) - 1)S^{\bold{k}}_{ia}G^{\bold{k}}_{a\sigma,b\sigma'}(\tau)S^{\bold{k}}_{bj}.
\end{align}

\section{Basis convergence of band energies \label{appendix:basis_conv_for_bands}}
\begin{figure}[tbh!]
\includegraphics[width=0.22\textwidth]{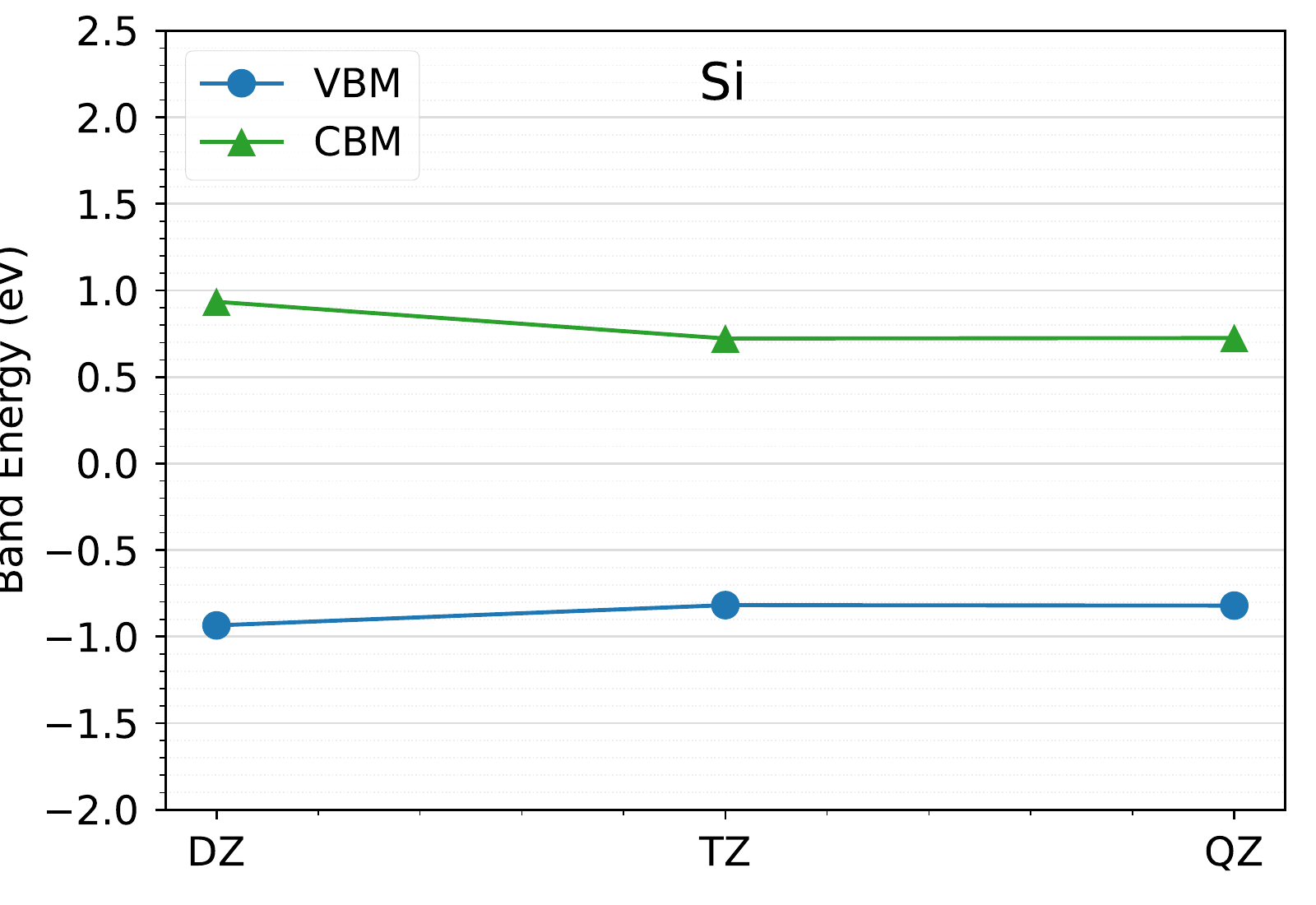}
\includegraphics[width=0.22\textwidth]{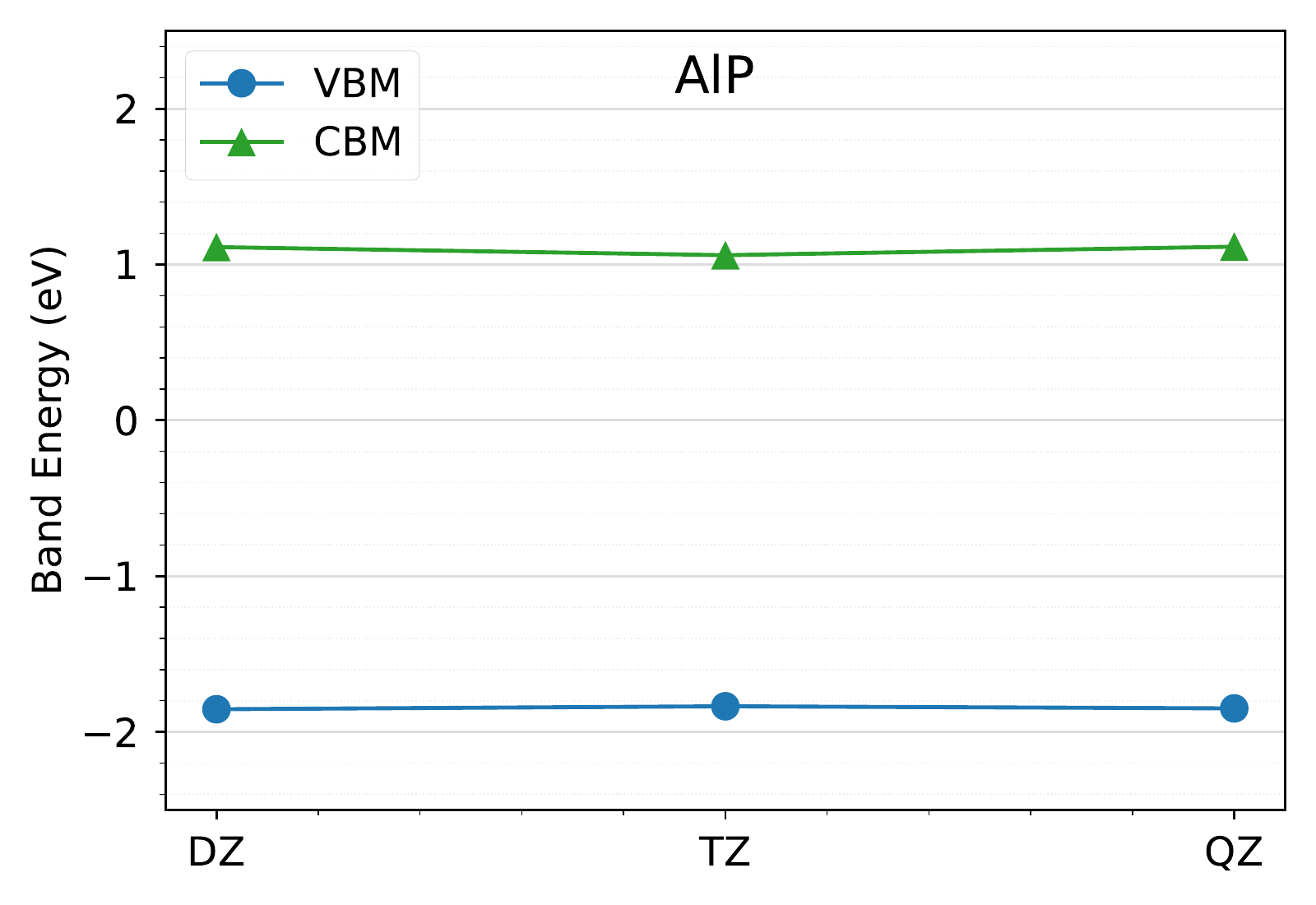} \\
\includegraphics[width=0.22\textwidth]{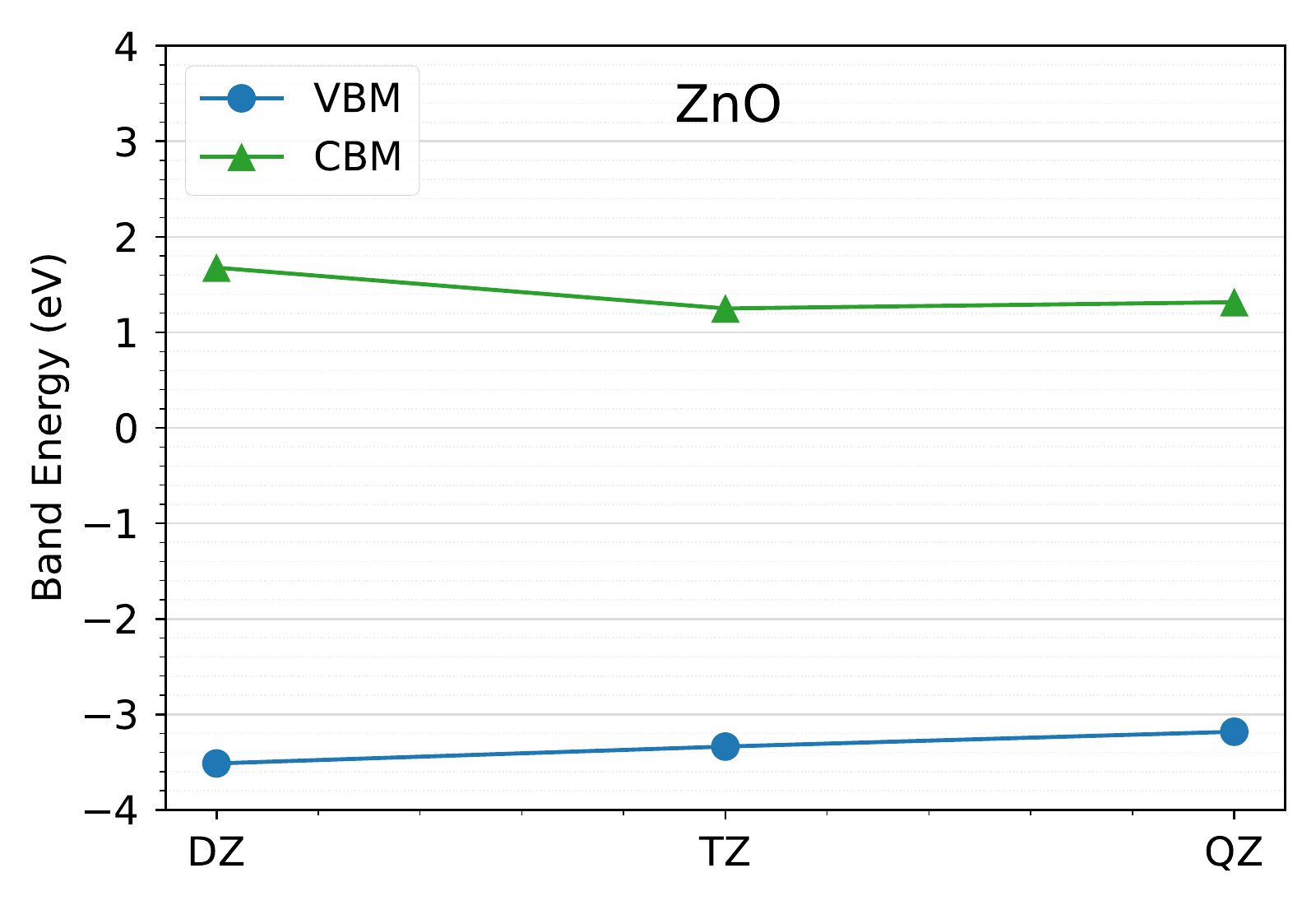}
\includegraphics[width=0.22\textwidth]{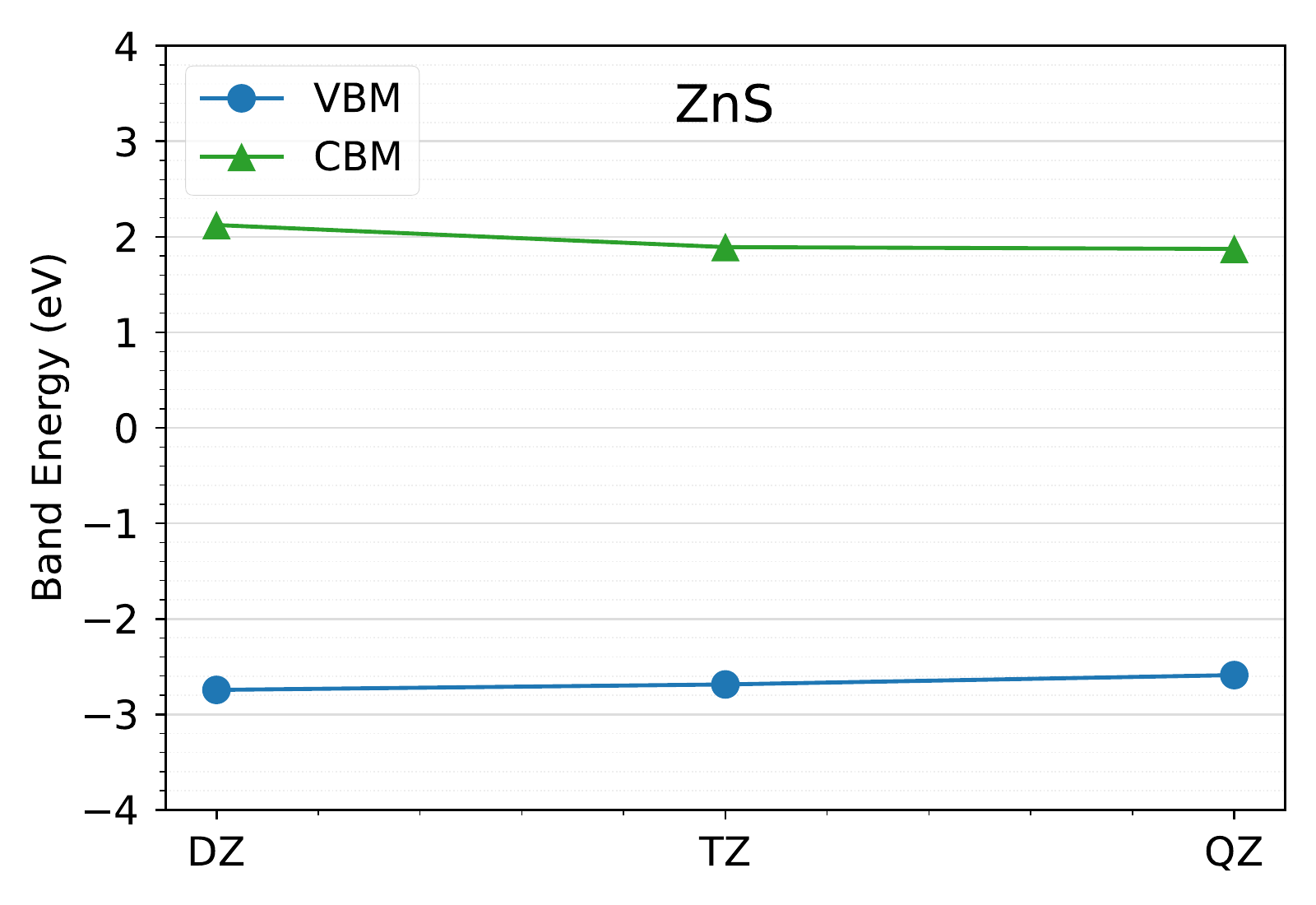}
\caption{The sc$GW$ valence band maximum (VBM) and the conduction band minimum (CBM) of Si, AlP, ZnO, and ZnS calculated using different basis sets. A $5\times5\times5$ $\bold{k}$-mesh is used for Si and AlP, and a $4\times4\times4$ $\bold{k}$-mesh is used for ZnO and ZnS. In x2c-QZVPAll, the most diffuse $s$ and $p$ functions of Si, Al, and Zn are removed to avoid linear dependencies.}\label{Fig:HOMO_LUMO_conv}
\end{figure}
Fig.~\ref{Fig:HOMO_LUMO_conv} shows the basis convergence of the valence band maximum (VBM) and the conduction band minimum (CBM) calculated using sc$GW$. 
Similar to Sec.~\ref{subsec:basis_conv}, the basis sets are systematically enlarged from x2c-SV(P)all (DZ), to x2c-TZVPall (TZ), and finally to x2c-QZVPall (QZ) basis set. 
Both VBM and CBM show similar convergence behavior compared to Table~\ref{tab:basis_conv}. 
This suggests that the basis set convergence of band gaps observed in Sec.~\ref{subsec:basis_conv} is not due to fortunate error cancellation. 

\end{document}